\documentclass{article}
\usepackage{amssymb,amsmath,graphicx}
\usepackage{amsfonts}
\usepackage{caption}
\usepackage{subfig}
\usepackage{epsfig,latexsym,graphicx}
\newtheorem{theorem}{Theorem}[section]

\newtheorem{corollary}[theorem]{Corollary}

\begin{document}

\label{firstpage}
\title{\bf{Robust Estimation in Generalised Linear Models : The Density Power Divergence Approach}
\thanks{This is part of the Ph.D. research work of the first author
 which is ongoing at the Indian Statistical Institute}
}
\author{Abhik Ghosh  and Ayanendranath Basu\\
Indian Statistical Institute \\
203 B. T. Road, Kolkata 700 108, India\\
{\it abhianik@gmail.com, ayanbasu@isical.ac.in}}
\date{}
\maketitle

\begin{abstract}
The generalised linear model (GLM) is a very important tool for analysing real data in biology, 
sociology, agriculture, engineering and many other application domain 
where the relationship between the response and explanatory variables
may not be linear or the distributions may not be normal in all the cases. 
However, quite often such real data contain a significant number of outliers in relation to the 
standard parametric model used in the analysis; in such cases the classical maximum likelihood 
estimator may fail to produce reasonable estimators and related inference could be unreliable.
In this paper, we develop a 
robust estimation procedure for the generalised linear models that can generate robust estimators
with little loss in efficiency. We will also explore two particular cases of the generalised linear model in details
--- Poisson regression for count data and logistic regression for binary data --- which are
widely applied in real life experiments. We will also illustrate the performance of the 
proposed estimators through several interesting data examples.
\end{abstract}


\section{Introduction}

Many real life problems require suitable techniques to describe some response data through a set of 
related explanatory variables. Parametric regression helps the experimenter to model 
such scenarios by means of some pre-specified functional relationship between response and 
explanatory variables described through a set of real parameters. 
The most widely used regression model is linear regression for 
continuous responses that depends on the covariates linearly. In practice, though, there are lots 
of different types of  response data like count data, binary response data and others which arise
frequently in real life experiments such as clinical trials, medical surveys, 
designed experiments etc. The generalised linear model is the general tool that can 
be used with all such types of response variables. This generalised linear model allows the experimenter
 to model the response variables by any distribution within a large family of distributions, 
 namely the exponential family, and the expected response by any (suitably smooth) function of the
 explanatory variables, with the only restriction that this function should depend on the 
 explanatory variables linearly. As a  special case, it also includes the ordinary linear 
 regression problem; the study of  the generalised linear model helps us to deal with a very large 
 superfamily of parametric regression problems.

The classical procedure to estimate the parameters of the generalised linear regression model is 
the maximum likelihood estimation method generating most efficient estimators. This theoretical 
advantage of the maximum likelihood estimator is, however, tempered by its 
known lack of robustness to outliers and model misspecification.
In many real life experiments, outliers show up as a matter or routine which influence the 
maximum likelihood estimators and often produce nonsensical results. 
So, there is a real need for developing robust estimation procedures for the 
  generalised linear regression model. Although there is a crowded field of robust estimators in the 
 ordinary  linear regression problem, there exist only a few for the generalised linear regression case. 
Cantoni and Ronchetti (2001) and Hosseinian (2009) present some such approaches; 
most of these approaches consider the explanatory variables to be stochastic. 
In this paper, we will develop an estimation procedure for the generalised linear model from 
a design perspective, where we will assume that the explanatory variables are fixed and given to us; 
each response is independent and follows the  same distribution specified by the generalised linear model, 
but having different distributional parameters depending on the values of the corresponding 
explanatory variables. 
The idea is motivated by the work of Ghosh and Basu (2013a) where a robust minimum divergence estimation 
procedure was developed under the general set-up of independent but non-homogeneous observations 
using the density power divergence. 
This work considered the case of simple linear regression in great detail and illustrated the 
properties of  the corresponding estimates of regression parameters demonstrating that many of them 
are highly robust with a very little loss in efficiency. Here, we will follow a similar approach 
to develop the minimum density power divergence estimators of the parameters of generalised linear model, 
which will be extremely robust in presence of the influential observations  
and also have comparable high efficiency.

The rest of the paper is organized as follows. In Section \ref{SEC:MDPDE_GLM} we will briefly describe
 the generalised regression model and develop the corresponding minimum density power estimators 
of parameters. We also prove the asymptotic properties and present the influence function analysis 
of the proposed minimum density power divergence estimator in case of generalised linear regression model
in this section. We will then explore the special case of Poisson regression for count data and 
logistic regression for binary data in Sections \ref{SEC:MDPDE_poisson} and \ref{SEC:MDPDE_logistic} 
respectively. Subsequently, we will present a short discussion on a data-driven choice of the 
tuning parameter $\alpha$ in Section \ref{SEC:alpha_GLM}. Section \ref{SEC:Data_examples} contains 
the application of the proposed minimum density power divergence estimation to some interesting 
real life data sets that illustrate the performance of the proposed methodology under several 
differing scenarios 
Finally the paper ends with some concluding remarks in Section \ref{SEC:conclusion}.

\section{The Minimum Density Power Divergence Estimator in Generalised Linear Models } \label{SEC:MDPDE_GLM}

\subsection{The Generalised Linear Model (GLM)}

	Generalised linear models are indeed generalizations of the normal linear regression model
where the response variables $Y_i$ are independent and assumed to follow  the general 
exponential family of distributions having density
\begin{equation}
f(y_i;\theta_i,\phi) = \exp\left\{ \frac{y_i\theta_i - b(\theta_i)}{a(\phi)} + c(y_i,\phi)  \right\},
\label{EQ:exp_family_density}
\end{equation}
where the canonical parameter $\theta_i$ is  a measure of location depending on 
the predictor $x_i$ and $\phi$ is the nuisance scale parameter.
The mean $\mu_i$ of $Y_i$ is linked to the explanatory variables $x_i$ through the relation 
$g(\mu_i) = \eta_i = x_i^T\beta,$
where $g$ is called the link function and $\eta_i = x_i^T\beta$ is known as the linear predictor.
Thus $g(\mu_i)$, rather than $\mu_i$ itself, is linearly related to the explanatory variables.
The link function $g$ is assumed to be monotone and differentiable. 
It is easy to see from the theory of exponential family of distributions that 
the canonical parameter $\theta_i$ is related to the mean $\mu_i$ by 
$\mu_i = b'(\theta_i)$
and also the variance of $y_i$ is given by 
$V(y_i) = b''(\theta_i)a(\phi),$
where $b'(\cdot)$ and $b''(\cdot)$ represent the first two derivatives of $b(\cdot)$ 
with respect to its argument. Thus our main parameter of interest in the generalised linear model becomes the 
regression coefficient $\beta$ and $\phi$ acts as the nuisance parameter which is involved 
only in the error variance. Clearly generalised linear model allows us to choose several possible densities 
$f$ from the exponential family including Normal, 
Binomial, Poisson, Exponential, Negative Binomial etc. and the link function $g$  to form a wide
 variety of regression models based on the  patterns of the available data. 
 Thus we can suitably model a large number of different types of data.

	Note that choosing $f$ to be the normal density and $g$ as the identity link function the 
generalised linear model reduces to  usual normal linear regression model. Further, choosing $f$ as the Poisson density 
and $g$ as the log link  $g(\mu) = \log(\mu)$, we get the Poisson regression case useful in modeling ordinal 
data and cases of over-dispersion. Also for the binomial $f$, choosing 
Logit link function $g(\mu)=\log(\mu(1-\mu))$ or Probit link function $g(\mu)=\Phi^{-1}(\mu)$, 
we get the  logistic and Probit regression models respectively which are useful in modeling the binary
 response variables. Similarly, many other useful models can be generated from the generalised linear model.

\subsection{The Minimum Density Power Divergence Estimator and its Estimating Equation}

	We will define the minimum density power divergence estimators for the generalised linear model with general density 
$f$ and link function $g$ so that we can estimate the regression coefficients for any regression model 
as a special case of it by substituting the form of $f$ and $g$. In the later 
sections, we will consider some of these special cases in detail. 
Since the form of the density power divergence is now well known in the literature we do not repeat 
it here.  A description of the form of this divergence may be found, for example, in Basu et al.~(1998) 
and Ghosh and Basu (2013a), among others.

	Let us assume that we have a data set $(y_i;x_i); i=1, \cdots, n$ from the 
generalised linear model with density $f$ given by equation (\ref{EQ:exp_family_density}) 
 and a general link function $g(\mu_i) = \eta_i = x_i^T\beta$. We will further assume that the 
independent variables $x_i$'s are given and fixed so that we are indeed considering a fixed carrier generalised linear model. 
Then we have the set-up of independent but non-homogeneous observations, where $y_1, \cdots, y_n$ 
are independent and $y_i$ has density  $f_i(.;(\beta, \phi)) = f(y_i;\theta_i,\phi)$ for all $i=1,\cdots, n$. 
Hence we can use the approach of Ghosh and Basu (2013a), where the minimum density power divergence 
estimator for the independent but non-homogeneous observations was defined. 
Following this approach, the minimum density power divergence estimator of $(\beta, \phi)$ 
has to be obtained by minimizing
$$
H_n(\beta, \phi) = \frac{1}{n} \sum_{i=1}^n V_i(Y_i;(\beta, \phi)),
$$
where
$$
V_i(Y_i;(\beta, \phi)) =  \int f_i(y;(\beta, \phi))^{1+\alpha} dy - 
\left(1+\frac{1}{\alpha}\right) f_i(Y_i;(\beta, \phi))^\alpha .
$$
Note that, in the usual generalised linear model estimation, we use a robust estimate of scale parameter $\phi$ 
and then estimate the regression parameter $\beta$. 
However, in the proposed minimum density power divergence estimation, 
we can simultaneously estimate $\beta$ and $\phi$ robustly by just minimizing 
$H_n(\beta, \phi)$ with respect to both the parameters. 
The estimating equation of the parameters are then given by
$\frac{1}{n} \sum_{i=1}^n \nabla V_i(Y_i;(\beta, \phi)) = 0$ or, 
\begin{eqnarray}
\frac{1+\alpha}{n} \sum_{i=1}^n \left[ \int u_i(y;(\beta, \phi)f_i(y;(\beta, \phi))^{1+\alpha}dy
 -  u_i(Y_i;(\beta, \phi) f_i(Y_i;(\beta, \phi))^\alpha \right] = 0.
\end{eqnarray}
where $u_i(y;(\beta, \phi))  = \nabla \log(f_i(y;(\beta, \phi))$; 
$\nabla$ represents the derivative with respect to $(\beta, \phi)$. 
Let $\nabla_\beta$ and $\nabla_\phi$ denote the individual derivatives with respect to 
$\beta$ and $\phi$ respectively. Then, a simple calculation shows that
\begin{equation}
\nabla_\beta \log(f_i(y_i;(\beta, \phi)) 
= \frac{(y_i - \mu_i)}{Var(y_i)g'(\mu_i)}  x_i = K_{1i}(y_i;(\beta, \phi)) x_i,
\end{equation}
and
\begin{equation}
\nabla_\phi \log(f_i(y_i;(\beta, \phi)) 
= -\frac{(y_i\theta_i - b(\theta_i))}{a^2(\phi)}a'(\phi) +\frac{\partial}{\partial \phi}c(y_i,\phi) 
= K_{2i}(y_i;(\beta, \phi)),
\end{equation}
where $K_{1i}$ and $K_{2i}$ are the indicated functions. Thus our estimating equations become
\begin{equation}
\sum_{i=1}^n x_i ~ \left[ \int K_{1i}(y;(\beta, \phi)) f_i(y;(\beta, \phi))^{1+\alpha}dy  - 
 K_{1i}(y_i;(\beta, \phi))  f_i(y_i;(\beta, \phi))^\alpha \right] = 0 ,
\label{EQ:est_eqn_beta}
\end{equation}
\begin{equation}
\sum_{i=1}^n \left[ \int K_{2i}(y;(\beta, \phi)) f_i(y;(\beta, \phi))^{1+\alpha}dy  - 
 K_{2i}(y_i;(\beta, \phi))  f_i(y_i;(\beta, \phi))^\alpha \right] = 0 .
 \label{EQ:est_eqn_phi}
\end{equation}
However, if we want to ignore the nuisance parameter $\phi$, as per the usual practice, 
and estimate $\beta$ taking $\phi$ fixed (or, substituted suitably), 
it is enough to consider only estimating equation (\ref{EQ:est_eqn_beta}). 
Further, for $\alpha=0$, we have
$\int \frac{(y_i - \mu_i)}{Var(y_i)} g'(\mu_i) x_i f_i(y_i;(\beta, \phi))^{1+\alpha}dy = 0$ 
and hence the estimating equations for $\beta$ (ignoring $\phi$) simplify to 
$$
\sum_{i=1}^n \frac{(Y_i - \mu_i)}{Var(Y_i)g'(\mu_i)}  x_i = 0.
$$
Note that this is just the maximum likelihood estimating equation and also is the same as
 the ordinary least squares (OLS) estimating equation for $\beta$ assuming $\phi$ to be fixed. 
 Thus the minimum density power divergence estimator  of $\beta$ with $\alpha=0$  equals 
 the maximum likelihood estimator as well as the ordinary least squares estimate of $\beta$. That is the minimum density power divergence estimator 
 proposed here is just a natural robust generalization of the maximum likelihood estimator.

	Also it is interesting to note that, if our density $f$ is such that 
$\int f(y;\theta_i,\phi)^{1+\alpha} dy$ is independent of the 
location parameter $\theta_i$, like the normal density, then we have \hfill\\\noindent
$\int \frac{(y_i - \mu_i)}{Var(y_i)g'(\mu_i)} x_i f_i(y_i;(\beta, \phi))^{1+\alpha}dy=0$ and 
hence the estimating equation (\ref{EQ:est_eqn_beta}) simplifies to 
\begin{equation}
\sum_{i=1}^n \frac{(Y_i - \mu_i)}{Var(Y_i)g'(\mu_i)} x_i  f_i(Y_i;(\beta, \phi))^\alpha  = 0 \label{EQ:est_eqn_beta1}.
\end{equation}


\subsection{Asymptotic Properties}\label{SEC:asymp_GLM}

	We will now derive the joint asymptotic distribution of the minimum density power divergence 
estimator $(\hat\beta, \hat\phi)$ of the parameters $(\beta, \phi)$ obtained by solving 
the estimating equations (\ref{EQ:est_eqn_beta}) and (\ref{EQ:est_eqn_phi}). 
For simplicity, we will assume that the true data generating distribution also belongs to 
the model density with parameters $(\beta^g, \phi^g)$.
Define, for $i=1, \ldots, n$, 
$$
\gamma_{1i} = \gamma_{1i}^{1+\alpha}(\beta, \phi) = \int K_{1i}(y;(\beta, \phi)) f_i(y;(\beta, \phi))^{1+\alpha}dy,
$$
$$
\gamma_{2i}=\gamma_{2i}^{1+\alpha}(\beta, \phi) = \int K_{2i}(y;(\beta, \phi)) f_i(y;(\beta, \phi))^{1+\alpha}dy,
$$
$$
\gamma_{jki} = \gamma_{jki}^{1+\alpha}(\beta, \phi) 
= \int K_{ji}(y;(\beta, \phi)) K_{ki}(y;(\beta, \phi)) f_i(y;(\beta, \phi))^{1+\alpha}dy,
~~ j, k = 1, 2,
$$
so that 
\begin{equation}
N_i^{1+\alpha}(\beta, \phi) = \int u_i(y;(\beta, \phi)f_i(y;(\beta, \phi))^{1+\alpha}dy = \left( \begin{array}{c}
\gamma_{1i} x_i \\
\gamma_{2i}
\end{array} \right),
\end{equation}
\begin{equation}
M_i^{1+\alpha}(\beta, \phi) = \int u_i(y;(\beta, \phi))u_i(y;(\beta, \phi))^T f_i(y;(\beta, \phi))^{1+\alpha}dy 
= \left( \begin{array}{c c}
\gamma_{11i} x_ix_i^T &  \gamma_{12i} x_i \\
\gamma_{12i} x_i^T & \gamma_{22i}
\end{array} \right).
\end{equation}
Now, put $\Gamma_{j}^{(\alpha)} = Diag(\gamma_{ji})_{i=1,\cdots,n}$ and 
$\Gamma_{jk}^{(\alpha)} = Diag(\gamma_{jki})_{i=1,\cdots,n}$ for $j,k=1,2$ and $X^T = [x_1, \cdots, x_n]$. 
Then we have 
\begin{equation}
\Psi_n(\beta, \phi) = \frac{1}{n}\sum_{i=1}^n M_i^{1+\alpha}(\beta, \phi) = \frac{1}{n} \left( \begin{array}{c c}
X^T\Gamma_{11}^{(\alpha)}X &  X^T\Gamma_{12}^{(\alpha)}\mathbf{1} \\
\mathbf{1}^T\Gamma_{12}^{(\alpha)}X & \mathbf{1}^T\Gamma_{22}^{(\alpha)}\mathbf{1}
\end{array} \right); 
\end{equation}
\begin{eqnarray}
 \Omega_n(\beta, \phi) &=& \frac{1}{n}\sum_{i=1}^n  \left[ M_i^{1+2\alpha}(\beta, \phi) - N_i^{1+\alpha}(\beta, \phi) (N_i^{1+\alpha}(\beta, \phi))^T \right] \\
 &=& \frac{1}{n} \left( \begin{array}{c c}
X^T[\Gamma_{11}^{(2\alpha)}-\Gamma_{1}^{(\alpha)T}\Gamma_{1}^{(\alpha)}]X &  X^T[\Gamma_{12}^{(2\alpha)}-\Gamma_{1}^{(\alpha)}\Gamma_{2}^{(\alpha)}]\mathbf{1} \\
\mathbf{1}^T[\Gamma_{12}^{(2\alpha)}-\Gamma_{1}^{(\alpha)}\Gamma_{2}^{(\alpha)}]X & \mathbf{1}^T[\Gamma_{22}^{(2\alpha)}-\Gamma_{2}^{(\alpha)T}\Gamma_{2}^{(\alpha)}]\mathbf{1}
\end{array} \right).
\end{eqnarray}
Then, the asymptotic distribution of $(\hat\beta, \hat\phi)$ follows from a simple modification of  
Theorem 3.1 of Ghosh and Basu (2013a), provided the Assumptions (A1) to (A7) hold in case of the generalised 
linear models. Note that, Assumptions (A1) to (A3) hold directly from the properties of 
the exponential family of distributions. 

\begin{theorem}\label{THM:asymp_GLM}
	Under Assumptions (A1)-(A7) of Ghosh and Basu (2013a), there exists a consistent sequence 
$(\hat\beta_n, \hat\phi_n)$  of roots to the minimum density power divergence estimating equations 
(\ref{EQ:est_eqn_beta}) and (\ref{EQ:est_eqn_phi}). Also, the asymptotic distribution of 
$\Omega_n^{-\frac{1}{2}}\Psi_n [\sqrt n ((\hat\beta_n, \hat\phi_n) - (\beta^g, \phi^g))]$ is
 $(p+1)-$dimensional normal with mean $0$ and variance $I_{p+1}$, the identity matrix of dimension $p+1$ 
 where $\Psi_n=\Psi_n(\beta^g, \phi^g)$ and $\Omega_n=\Omega_n(\beta^g, \phi^g)$.\\
\end{theorem}

	It follows from above theorem that the reciprocal of the matrix $\Psi_n^{-1}\Omega_n\Psi_n^{-1}$ 
gives a estimate of the asymptotic efficiency of the minimum density power divergence estimators 
$(\hat\beta_n, \hat\phi_n)$. Though this depends on the sample size $n$ and the given covariates 
$x_i$'s, it will give a reasonable estimate of the asymptotic efficiency for large $n$.
We will examine the performance of this measure for several particular problems in the subsequent sections.

Further, note that the asymptotic covariance of the estimators $\hat\beta_n$ and $\hat\phi_n$ 
are not in general $0$ and hence these estimators are not asymptotically independent for all 
the generalised regression models. However, for some particular cases including the 
normal linear regression case, they turn out to be independent. One possible set of sufficient conditions for 
their independence are $\gamma_{12i}^{1+2\alpha}=0$ and $\gamma_{1i}^{1+\alpha} \gamma_{2i}^{1+\alpha} =0$ 
for all $i$. These conditions hold for the normal linear regression case.

\subsection{Influence Function}\label{SEC:IF_GLM}

	To illustrate the robustness properties of the proposed estimation methodology for the 
generalised regression model, we will now consider the influence function of the minimum 
density power divergence estimator of the parameter $\theta = (\beta, \phi)$. 
For this we need to consider them in terms of a statistical functional 
at the true data generating distributions. Let $T_\alpha^\beta(G_1,\cdots,G_n)$  and 
$T_\alpha^\phi(G_1,\cdots,G_n)$  denote the minimum density  power divergence functional 
for the parameters $\beta$ and $\phi$ respectively. 
Let $T_\alpha(G_1,\cdots,G_n) = ( T_\alpha^\beta(G_1,\cdots,G_n)^T , T_\alpha^\phi(G_1,\cdots,G_n))^T $,
 which is defined by  
$$
\frac{1}{n} \sum_{i=1}^n d_\alpha(g_i(.),f_i(.;T_\alpha(G_1,\cdots,G_n))) = \min_{\theta \in \Theta}
~  \frac{1}{n} \sum_{i=1}^n d_\alpha(g_i(.),f_i(.;\theta)),
$$
where $g_i$ is the probability density function corresponding to $G_i$.
We consider the contaminated density $g_{i,\epsilon} = (1-\epsilon) g_i + \epsilon \delta_{t_i}$ 
where $t_i$ is the point of contamination and $G_{i,\epsilon}$ denotes the corresponding distribution function for 
all $i=1, \cdots, n$. Let $\theta_\epsilon^{i_0} = T_\alpha(G_1,\cdot,G_{i_0-1},G_{i_0,\epsilon},\cdot, G_n)$
 be the minimum density power divergence functional with contamination only in the $i_0^{th}$ direction. 
Then a fairly straightforward (albeit lengthy and tedious) calculation shows that 
the influence function of $T_\alpha$ for contamination at the direction $i_0$ will be
\begin{eqnarray}
IF_{i_0}(t_{i_0}, T_\alpha, G_1, \cdots, G_n) 
&=& \Psi_n^{-1} \frac{1}{n}\left[  f_{i_0}(t_{i_0};(\beta, \phi))^\alpha u_{i_0}(t_{i_0};(\beta, \phi)) 
-  N_{i_0}^{1+\alpha} \right]\\
 &=& \Psi_n^{-1} \frac{1}{n} \left( \begin{array}{c}
\left[f_{i_0}(t_{i_0};(\beta, \phi))^\alpha  K_{1i_0}(t_{i_0};(\beta, \phi)) - \gamma_{1i_0}\right] x_i \\
\\
f_{i_0}(t_{i_0};(\beta, \phi))^\alpha  K_{2i_0}(t_{i_0};(\beta, \phi)) - \gamma_{2i_0}
\end{array} \right).
\end{eqnarray}
Note that for any fixed sample size $n$ and any given (finite) values of $X_i$'s, 
if $\Psi_n$ and $\gamma_{ji_0}$'s are assumed to be bounded,  the influence function of 
the minimum density power divergence estimator of the parameters $(\beta, \phi)$ will be bounded with respect to the contamination in any 
direction $i_0$ provided the terms $f_{i}(t_{i};(\beta, \phi))^\alpha  K_{ji}(t_{i};(\beta, \phi))$ 
are bounded for all $i$ and $j=1,2$. Under assumptions (A1) to (A7) of Ghosh and Basu (2013a), 
the values of $\Psi_n$ and $\gamma_{ji_0}$'s are necessarily bounded. 
This can be easily seen to hold for the majority of 
generalised linear models with $\alpha>0$ because of the exponential nature of the density function 
and the polynomial nature of the functions $ K_{ji}(t_{i};(\beta, \phi))$. 
This demonstrates the robust nature of the minimum density power divergence estimator in most generalised linear models with $\alpha>0$. However, for $\alpha=0$ 
the term   $f_{i}(t_{i};(\beta, \phi))^\alpha  K_{ji}(t_{i};(\beta, \phi)) = K_{ji}(t_{i};(\beta, \phi))$ 
is clearly unbounded implying the non-robust nature of the maximum likelihood estimator in the case of any generalised linear model.

As in Ghosh and Basu (2013a), in this context also we can define some measures of sensitivity 
based on the influence function. The (unstandardized) \textbf{gross-error sensitivity}
 of the functional $T_\alpha$ at the true distributions $G_1,\cdots,G_n$, 
 considering contamination only in the $i_0^{th}$ direction may be  defined as
\begin{eqnarray}
&\gamma_{i_0}^u (T_\alpha, G_1, \cdots, G_n) = \sup_{t_{i_0}} \{||IF_{i_0}(_{i_0}, T_\alpha, G_1, \cdots, G_n)||\} 
\label{EQ:gamma_u_1}\\
 \nonumber\\
& ~~~~ =  \frac{1}{n} \sup_{t_{i_0}} \{\left[  f_{i_0}(t_{i_0};\theta)^\alpha u_{i_0}(t_{i_0};\theta) 
- \xi_{i_0} \right]^T \Psi_n^{-2} 
 \left[  f_{i_0}(t_{i_0};\theta)^\alpha u_{i_0}(t_{i_0};\theta) - \xi_{i_0} \right] \}^{\frac{1}{2}}.
\label{EQ:gamma_u_2}
\end{eqnarray}
 Since it is not invariant to scale transformations of the individual parameter components, we will consider 
the \textbf{Self-Standardized Sensitivity}. For contamination in only the $i_0^{th}$ direction, 
this may be defined as 

\begin{eqnarray}
&&\gamma_{i_0}^s (T_\alpha, G_1, \cdots, G_n) \nonumber \\
&& ~~~~ = \sup_{t_{i_0}} \{IF_{i_0}(t_{i_0}, T_\alpha, G_1, \cdots, G_n)^T 
(\Psi_n^{-1}\Omega_n \Psi_n^{-1})^{-1}IF_{i_0}(t_{i_0}, T_\alpha, G_1, \cdots, G_n)\}^{\frac{1}{2}}, 
~~~~ \label{EQ:gamma_s_1} \\
 \nonumber\\
&& ~~~~ = \frac{1}{n} \sup_{t_{i_0}} \{\left[  f_{i_0}(t_{i_0};\theta)^\alpha u_{i_0}(t_{i_0};\theta) 
- \xi_{i_0} \right]^T \Omega_n^{-1}
\left[  f_{i_0}(t_{i_0};\theta)^\alpha u_{i_0}(t_{i_0};\theta) - \xi_{i_0} \right] \}^{\frac{1}{2}}. 
\label{EQ:gamma_s_2}
\end{eqnarray}

	As discussed before, in the particular case when $\gamma_{12i}^{1+2\alpha}=0$ and 
$\gamma_{1i}^{1+\alpha} \gamma_{2i}^{1+\alpha} =0$ for all $i$ (like the normal linear 	regression case), 
the minimum density power divergence estimator of $\beta$ and $\phi$ become asymptotically independent and we can also separate out 
the influence function for the minimum density power divergence estimator of $\beta$ and $\phi$. Due to the special form of 
the matrix $\Psi_n$ in this case, these two influence functions simplify to 
$$
IF_{i_0}(t_{i_0}, T_\alpha^\beta, G_1, \cdots, G_n) =  (X^T\Gamma_{11}^{(\alpha)}X)^{-1} x_{i_0} 
\left[f_{i_0}(t_{i_0};(\beta, \phi))^\alpha  K_{1i_0}(t_{i_0};(\beta, \phi)) - \gamma_{1i_0}\right],
$$
and
$$
IF_{i_0}(t_{i_0}, T_\alpha^\phi, G_1, \cdots, G_n) =  (\mathbf{1}^T\Gamma_{22}^{(\alpha)}\mathbf{1})^{-1} 
\left[f_{i_0}(t_{i_0};(\beta, \phi))^\alpha  K_{2i_0}(t_{i_0};(\beta, \phi)) - \gamma_{2i_0}\right] 
$$
respectively. Then the (unstandardized) gross-error sensitivities of  $\beta$ and $\phi$ are  given 
respectively by 
$$
\gamma_{i_0}^u (T_\alpha^\beta, G_1, \cdots, G_n) 
=  \sqrt{x_{i_0}^T(X^T\Gamma_{11}^{(\alpha)}X)^{-1}x_{i_0}} \sup_{t_{i_0}} ~ 
\left|f_{i_0}(t_{i_0};(\beta, \phi))^\alpha  K_{1i_0}(t_{i_0};(\beta, \phi)) - \gamma_{1i_0}\right|,
$$
and
$$
\gamma_{i_0}^u (T_\alpha^\phi, G_1, \cdots, G_n) =  (\mathbf{1}^T\Gamma_{22}^{(\alpha)}\mathbf{1})^{-1/2} 
\left|f_{i_0}(t_{i_0};(\beta, \phi))^\alpha  K_{2i_0}(t_{i_0};(\beta, \phi)) - \gamma_{2i_0}\right|.
$$
The corresponding Self-Standardized Sensitivities also have the same form as above with 
$\Gamma_{11}^{(\alpha)}$ and $\Gamma_{22}^{(\alpha)}$ replaced by 
$[\Gamma_{11}^{(2\alpha)}-\Gamma_{1}^{(\alpha)2}]$ and $[\Gamma_{22}^{(2\alpha)}-\Gamma_{2}^{(\alpha)2}]$ 
respectively. Clearly these sensitivities are infinite for $\alpha=0$ and are generally finite 
for most of the generalised linear model with $\alpha>0$.

In the following we will derive the explicit form of the influence functions and 
sensitivity measures for some special cases of the generalised linear model.

\section{Special Case (I) : Poisson Regression for Count Data}\label{SEC:MDPDE_poisson}

	Samples that have the count data structure are usually modeled by the 
 Poisson distribution. And the most useful regression tool for count data is the 
Poisson regression model where, given the values of explanatory variables, dependent variables 
independently follow the  Poisson distribution but with different mean parameters depending on the 
corresponding values of the explanatory variable. More precisely, let $(y_1,x_1), \cdots, (y_n, x_n)$ be 
the sample observations from the Poisson regression model. We will assume that the values $x_i$  
of the explanatory  variable are fixed known values. Then, in the Poisson regression model, the count variable 
$y_i$s are assumed to be independent and  have Poisson distributions with
$$
E(y_i|x_i) = e^{(x_i^T\beta)}
$$
and we want to estimate the parameter $\beta$ efficiently and robustly.

\subsection{The minimum density power divergence estimator for Poisson Regression}

Poisson regression is indeed a special case of generalised linear models with known shape parameter $\phi=1$ 
and $\theta_i = \eta_i = x_i^T\beta$, $b(\theta_i) = e^{\theta_i}$ and $c(y_i) = - \log(y_i !)$. 
Since here the mean is 
$\mu_i = e^{(x_i^T\beta)}=e^{\eta_i}$, the link function $g$ is the natural logarithm function and 
the variance of $y_i$ is also $e^{(x_i^T\beta)}$. Thus, we can estimate the unknown parameter $\beta$ using
 our minimum density power divergence estimation procedure as described earlier. 
Using the above notation and the form of the Poisson distribution, we get
\begin{eqnarray}
N_i^{1+\alpha}(\beta) &=& \sum_{y=0}^\infty \left(y- e^{(x_i^T\beta)}\right)x_i f_i(y;\beta)^{1+\alpha}
=\gamma_{1i}x_i,\\
M_i^{1+\alpha}(\beta) &=& \sum_{y=0}^\infty \left(y- e^{(x_i^T\beta)}\right)^2 (x_i x_i^T) f_i(y;\beta)^{1+\alpha}
=\gamma_{11i} (x_i x_i^T) ,
\end{eqnarray}
where $f_i(y;\beta)$ is the probability mass function of the Poisson distribution with mean $e^{(x_i^T\beta)}$. 
Then for $\alpha \geq 0$, the minimum density power divergence estimating equation is given by 
\begin{equation}
\sum_{i=1}^n \left[ \gamma_{1i}(\beta) - \left(y_i- e^{(x_i^T\beta)}\right)f_i(y;\beta)^{\alpha} \right]x_i  = 0. 
\label{EQ:est_eqn_PoissReg}
\end{equation}
In particular, for $\alpha=0$ above estimating equation simplifies to the maximum likelihood 
estimating equation given by
\begin{equation}
\sum_{i=1}^n \left(y_i- e^{(x_i^T\beta)}\right)x_i = 0. 
\end{equation}
However, for $\alpha > 0$, there is no simplified form for $\gamma_{1i}$  and $\gamma_{11i} $ so that we 
need to compute this quantities numerically and then numerically solve the estimating equation 
(\ref{EQ:est_eqn_PoissReg}) with respect to $\beta$.

\subsection{Properties of the minimum density power divergence estimator}\label{SEC:asymp_poisson}

	 The asymptotic properties of the minimum density power divergence estimator of $\beta$ in this special case follows directly from the
Theorem \ref{THM:asymp_GLM} of the previous section. Under the notation of Section \ref{SEC:asymp_GLM} 
we have 
$$
\Psi_n(\beta) = \frac{1}{n} \left(X^T\Gamma_{11}^{(\alpha)}X\right), ~~~~~~
 \Omega_n(\beta) = \frac{1}{n} \left(X^T[\Gamma_{11}^{(2\alpha)}-\Gamma_{1}^{(\alpha)2}]X\right).
$$
Thus we have
\begin{corollary}
Under Assumptions (A1)-(A7) of Ghosh and Basu (2013a), there exists a consistent sequence 
$\hat\beta_n=\hat\beta_n^{(\alpha)}$ of roots to the minimum density power divergence estimating equations (\ref{EQ:est_eqn_PoissReg})
at tuning parameter $\alpha$. Also, the asymptotic distribution of 
$\left(X^T[\Gamma_{11}^{(2\alpha)}(\beta^g)-\Gamma_{1}^{(\alpha)2}(\beta^g)]X\right)^{-\frac{1}{2}}
\left(X^T\Gamma_{11}^{(\alpha)}(\beta^g)X\right)(\hat\beta_n - \beta^g)$ is
 $p-$dimensional normal with mean $0$ and variance $I_{p}$.
\end{corollary}

Thus the asymptotic efficiency of the different minimum density power divergence estimator $\hat\beta_n = \hat\beta_n^{(\alpha)}$ of $\beta$ 
can be measured based on the asymptotic variance 
$$ AV_\alpha(\beta^g) = \left(X^T\Gamma_{11}^{(\alpha)}(\beta^g)X\right)^{-1} 
\left(X^T[\Gamma_{11}^{(2\alpha)}(\beta^g)-\Gamma_{1}^{(\alpha)2}(\beta^g)]X\right) 
\left(X^T\Gamma_{11}^{(\alpha)}(\beta^g)X\right)^{-1},
$$
 which can be consistently estimated by replacing $\beta^g$ with $\hat{\beta_n}$ in its expression,
i.e., $\widehat{AV}_\alpha = AV_\alpha(\hat\beta_n)$. Thus an estimate of the relative efficiency 
of the different minimum density power divergence estimators of the $i^{\mbox{th}}$ component of the parameter vector $\beta$ with respect 
to its maximum likelihood estimator (or the ordinary least squares estimator) is given by 
$$
\widehat{RE}_{i,\alpha} = \frac{i^{\mbox{th}} \mbox{ diagonal entry of } \widehat{AV}_0}{i^{\mbox{th}} 
\mbox{ diagonal entry of } \widehat{AV}_\alpha} \times 100.
$$

Clearly, the above estimate of the relative efficiency depends on the sample size $n$ and 
the choice of the given explanatory variables $x_i$'s. But it can be shown that the consistency 
of the estimator $\hat{\beta_n}$ implies that the above measure gives us a consistent estimator 
of the asymptotic relative efficiency if the $x_i$'s are chosen suitably. 
For example we must have $X^TX$ to be bounded. We  have presented the empirical value of 
this measure of relative efficiency for different sample sizes  $n=50$ and $n=100$ respectively 
under several different cases in Tables \ref{TAB:ARE_poiss_50} and \ref{TAB:ARE_poiss_100}.
We have reported six cases which are defined based on the true 
values of the regression coefficients $\beta=(\beta_0, \beta_1, \ldots, \beta_p)$ and 
the given values of the explanatory variables $x_i$ ($i=1,\ldots,n$) as follows:
\begin{eqnarray}
\mbox{Case I~~} &:& p=2; \beta=(1, 1) \mbox{  and  } x_i = (1, \sqrt{i}). \nonumber\\
\mbox{Case II~} &:& p=2; \beta=(1, 0.5) \mbox{ and } x_i = (1, \sqrt{i}). \nonumber\\
\mbox{Case III} &:&  p=2; \beta=(1, 1) \mbox{  and  } x_i = (1, \frac{1}{i}). \nonumber\\
\mbox{Case IV} &:&  p=2; \beta=(1, 0.5) \mbox{  and  } x_i = (1, \frac{1}{i}). \nonumber\\
\mbox{Case V~} &:&  p=3; \beta=(1, 1, 1) \mbox{  and  } x_i = (1, \sqrt{i}, \frac{1}{i^2}).\nonumber\\
\mbox{Case VI} &:&  p=3; \beta=(2, 1, 0.5) \mbox{  and  } x_i = (1, \sqrt{i}, \frac{1}{i^2}).\nonumber
\end{eqnarray}
All the simulations are done based on $1000$ replications. 
It is clear from the tables that the loss of efficiency is very negligible for 
the minimum density power divergence estimator with small positive $\alpha$ 
under each of the cases considered here. 
Even for large positive $\alpha$ near $0.5$ also, we can get quite high efficiency if $x_i$s are relatively small.

Next, in order to see the robustness of the minimum density power divergence estimator in case of the Poisson regression model, 
 we will use the results from the Section \ref{SEC:IF_GLM}. 
 The influence function of the minimum density power divergence estimator in the direction $i_0$ simplifies to
$$
IF_{i_0}(t_{i_0}, T_\alpha^\beta, G_1, \cdots, G_n) =  (X^T\Gamma_{11}^{(\alpha)}X)^{-1} x_{i_0} 
\left[\left(t_{i_0} - e^{(x_{i_0}^T\beta)}\right) \frac{e^{\alpha t_{i_0}(x_{i_0}^T\beta)}}{(t_{i_0}!)^\alpha} 
e^{\alpha e^{(x_{i_0}^T\beta)}} - \gamma_{1i_0}\right].
$$

\begin{table}[h]
\caption{The estimated Relative efficiencies of the MDPDE for various values of the tuning parameter $\alpha$  under different cases of Poisson regression with sample size $n = 50$}
\label{TAB:ARE_poiss_50}
\begin{center}
\resizebox{0.8\textwidth}{!}{
\begin{tabular}{llrrrrrrrr} \hline
Case	& Coefficients	&	$\alpha = 0$	&	$\alpha = 0.01$	&	$\alpha = 0.1$	&	$\alpha = 0.25$	&	$\alpha = 0.4$	&	$\alpha = 0.5$	&	$\alpha = 0.7$	&	$\alpha = 1$
	\\	 \hline
I	&	$\beta_0$	&	100.0	&	100.0	&	98.3	&	91.2	&	80.9	&	73.5	&	58.1	&	37.9	\\	
	&	$\beta_1$	&	100.0	&	100.0	&	98.3	&	91.1	&	80.7	&	73.1	&	57.2	&	36.4	\\	\hline
II	&	$\beta_0$	&	100.0	&	99.9	&	98.5	&	93.2	&	85.9	&	80.7	&	70.5	&	56.5	\\	 
	&	$\beta_1$	&	100.0	&	99.8	&	98.4	&	93.0	&	85.7	&	80.5	&	70.0	&	55.5	\\	\hline
III	&	$\beta_0$	&	100.0	&	100.0	&	98.8	&	94.5	&	88.9	&	85.1	&	77.6	&	67.7	\\	 
	&	$\beta_1$	&	100.0	&	100.0	&	98.8	&	93.7	&	88.4	&	84.3	&	76.0	&	64.8	\\	\hline
IV	&	$\beta_0$	&	100.0	&	100.0	&	98.7	&	94.4	&	88.9	&	85.1	&	77.8	&	68.0	\\	 
	&	$\beta_1$	&	100.0	&	100.0	&	98.9	&	94.3	&	88.4	&	84.6	&	76.6	&	66.3	\\	\hline
V &	$\beta_0$	&	100.0	&	100.0	&	98.9	&	94.4	&	88.7	&	84.9	&	77.4	&	67.5	\\	 
	&	$\beta_1$	&	100.0	&	100.0	&	98.9	&	94.3	&	88.1	&	84.3	&	76.1	&	65.7	\\	\hline
	&	$\beta_2$	&	100.0	&	100.0	&	98.9	&	94.1	&	88.1	&	84.2	&	75.9	&	65.4	\\	 
VI	&	$\beta_0$	&	100.0	&	100.0	&	98.7	&	94.2	&	88.1	&	84.0	&	76.0	&	65.5	\\	\hline
	&	$\beta_1$	&	100.0	&	100.0	&	98.6	&	94.3	&	88.1	&	83.8	&	75.8	&	65.0	\\	 
	&	$\beta_2$	&	100.0	&	100.0	&	98.6	&	94.3	&	88.1	&	83.7	&	75.7	&	64.8\\ \hline
\end{tabular}}
\end{center}
\end{table}

\begin{table}[!t] 
\caption{The estimated Relative efficiencies of the MDPDE for various values of the tuning parameter $\alpha$  under different cases of Poisson regression with sample size $n = 100$}
\label{TAB:ARE_poiss_100}
\begin{center}
\resizebox{0.8\textwidth}{!}{
\begin{tabular}{llrrrrrrrr} \hline
Case	& Coefficients	&	$\alpha = 0$	&	$\alpha = 0.01$	&	$\alpha = 0.1$	&	$\alpha = 0.25$	&	$\alpha = 0.4$	&	$\alpha = 0.5$	&	$\alpha = 0.7$	&	$\alpha = 1$
	\\	 \hline
I	&	$\beta_0$	&	100.0	&	100.0	&	98.2	&	89.8	&	77.3	&	67.7	&	48.0	&	24.1	\\	
	&	$\beta_1$	&	100.0	&	100.0	&	98.2	&	89.7	&	77.0	&	67.2	&	47.1	&	22.9	\\	\hline
II	&	$\beta_0$	&	100.0	&	100.0	&	98.4	&	92.4	&	83.9	&	77.9	&	65.4	&	48.5	\\	 
	&	$\beta_1$	&	100.0	&	100.0	&	98.4	&	92.3	&	83.7	&	77.5	&	64.7	&	47.2	\\	\hline
III	&	$\beta_0$	&	100.0	&	100.0	&	98.7	&	94.4	&	88.9	&	85.1	&	77.8	&	67.9	\\	 
	&	$\beta_1$	&	100.0	&	100.0	&	98.8	&	94.4	&	88.3	&	83.9	&	75.6	&	64.8	\\	\hline
IV	&	$\beta_0$	&	100.0	&	100.0	&	98.9	&	94.4	&	89.0	&	85.2	&	78.0	&	68.2	\\	 
	&	$\beta_1$	&	100.0	&	100.0	&	99.4	&	93.8	&	89.0	&	84.8	&	76.9	&	66.7	\\	\hline
V	&	$\beta_0$	&	100.0	&	100.0	&	98.7	&	94.3	&	88.9	&	85.0	&	77.7	&	67.7	\\	 
	&	$\beta_1$	&	100.0	&	99.9	&	98.6	&	93.8	&	88.2	&	83.9	&	76.2	&	65.6	\\	\hline
	&	$\beta_2$	&	100.0	&	99.9	&	98.6	&	93.8	&	88.2	&	83.8	&	76.0	&	65.2	\\	 
VI	&	$\beta_0$	&	100.0	&	100.0	&	98.9	&	94.2	&	88.2	&	84.1	&	76.2	&	65.6	\\	\hline
	&	$\beta_1$	&	100.0	&	100.0	&	99.2	&	94.2	&	88.3	&	84.2	&	76.0	&	65.0	\\	 
	&	$\beta_2$	&	100.0	&	100.0	&	99.1	&	94.2	&	88.3	&	84.1	&	75.7	&	64.8\\\hline
\end{tabular}}
\end{center}
\end{table}

\newpage
 
\begin{figure}[h]
\centering
\subfloat[Model I, $i_0$ = 1]{
\includegraphics[width=0.5\textwidth, height=0.35\textwidth]{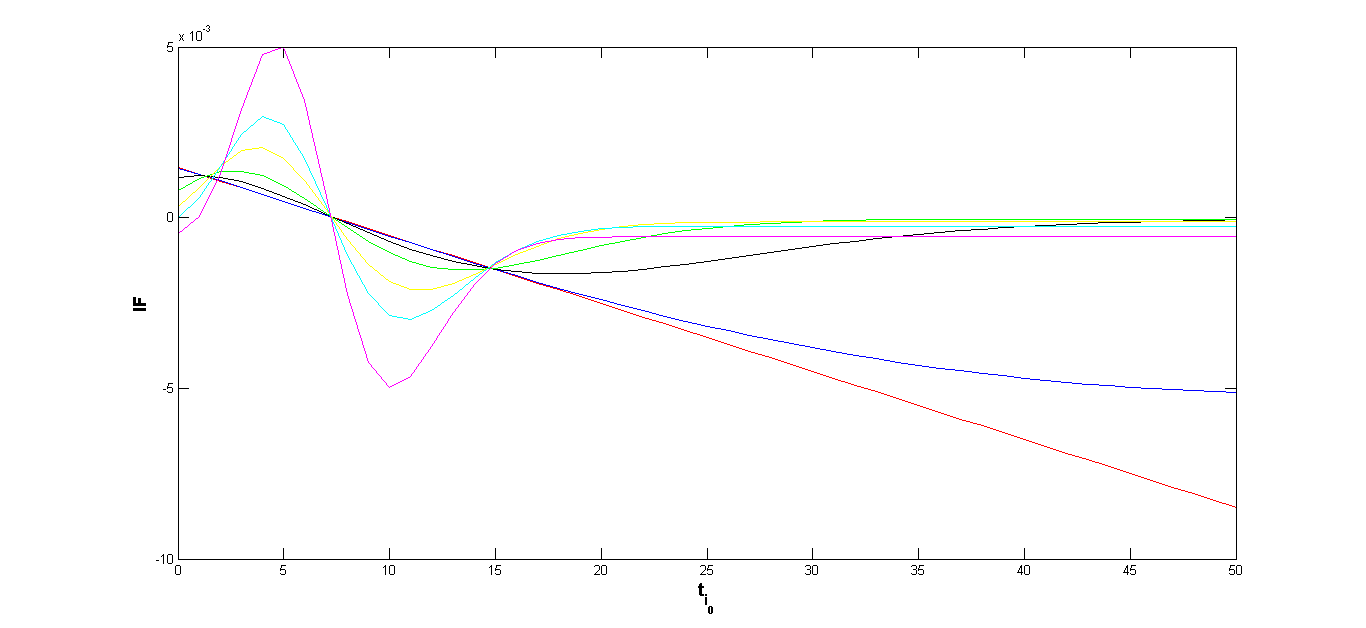}
\label{fig:11_50}}
~ 
\subfloat[Model I, $i_0$ = 20]{
\includegraphics[width=0.5\textwidth, height=0.35\textwidth]{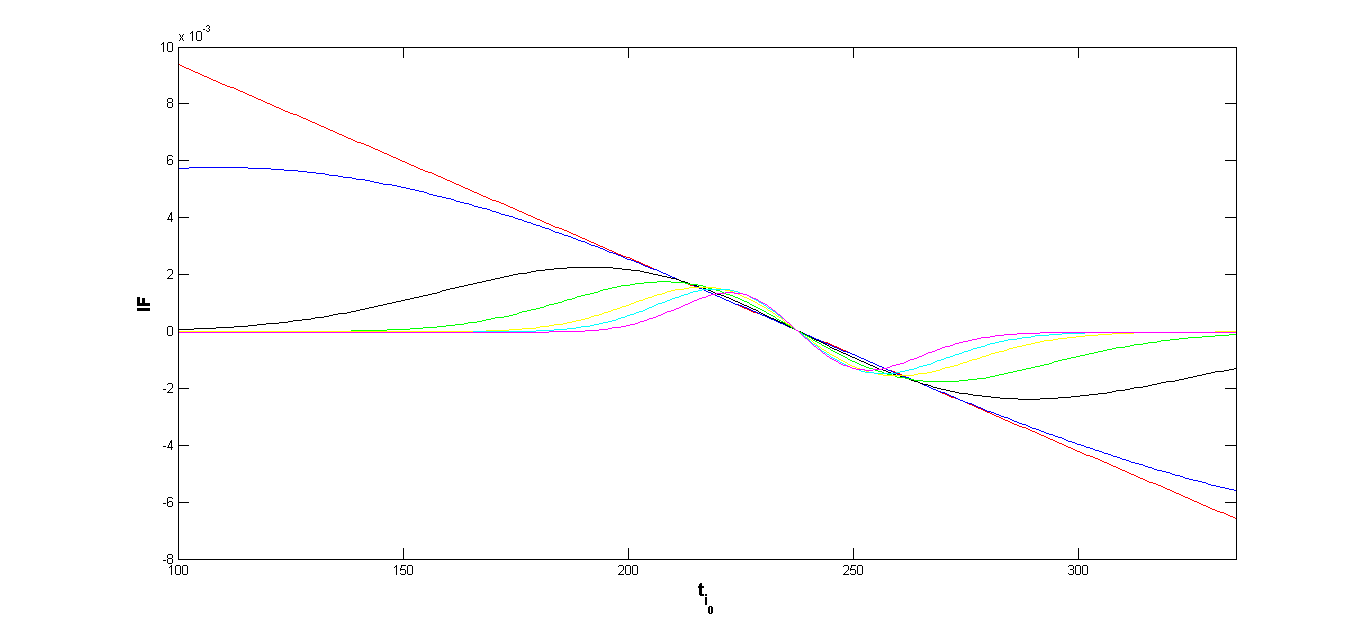}
\label{fig:12_50}}
\\
\subfloat[Model II, $i_0$ = 1]{
\includegraphics[width=0.5\textwidth, height=0.35\textwidth]{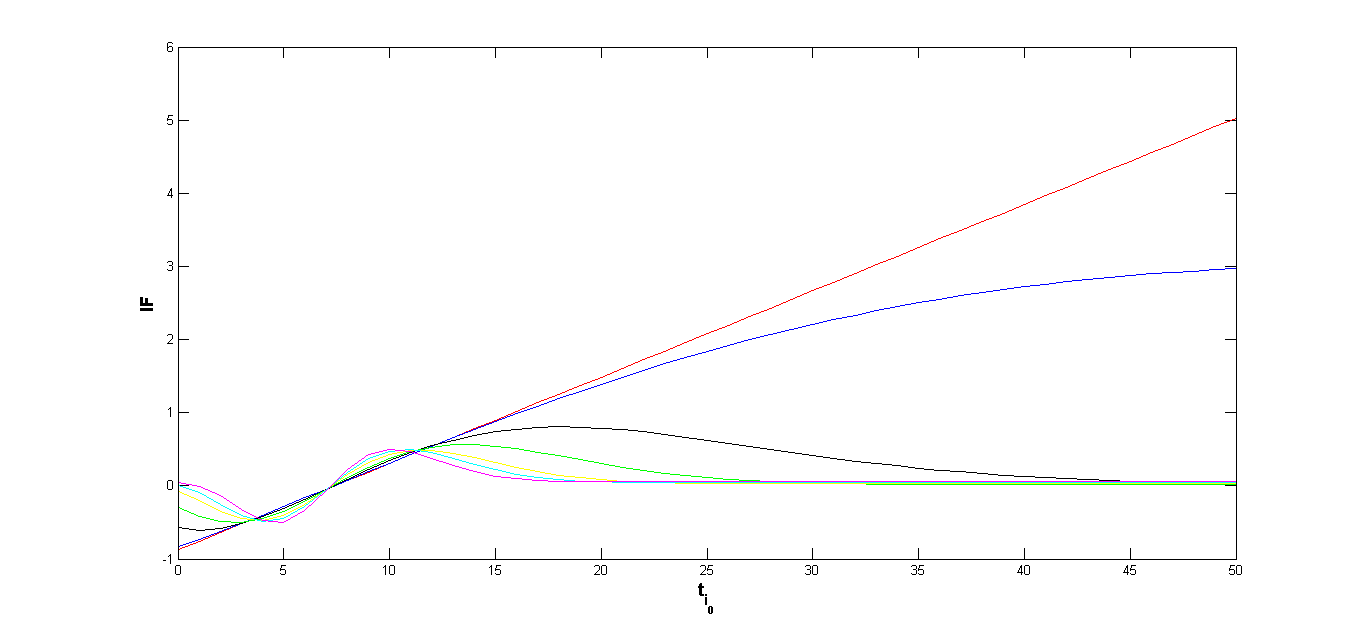}
\label{fig:21_50}}
~ 
\subfloat[Model II, $i_0$ = 50]{
\includegraphics[width=0.5\textwidth, height=0.35\textwidth]{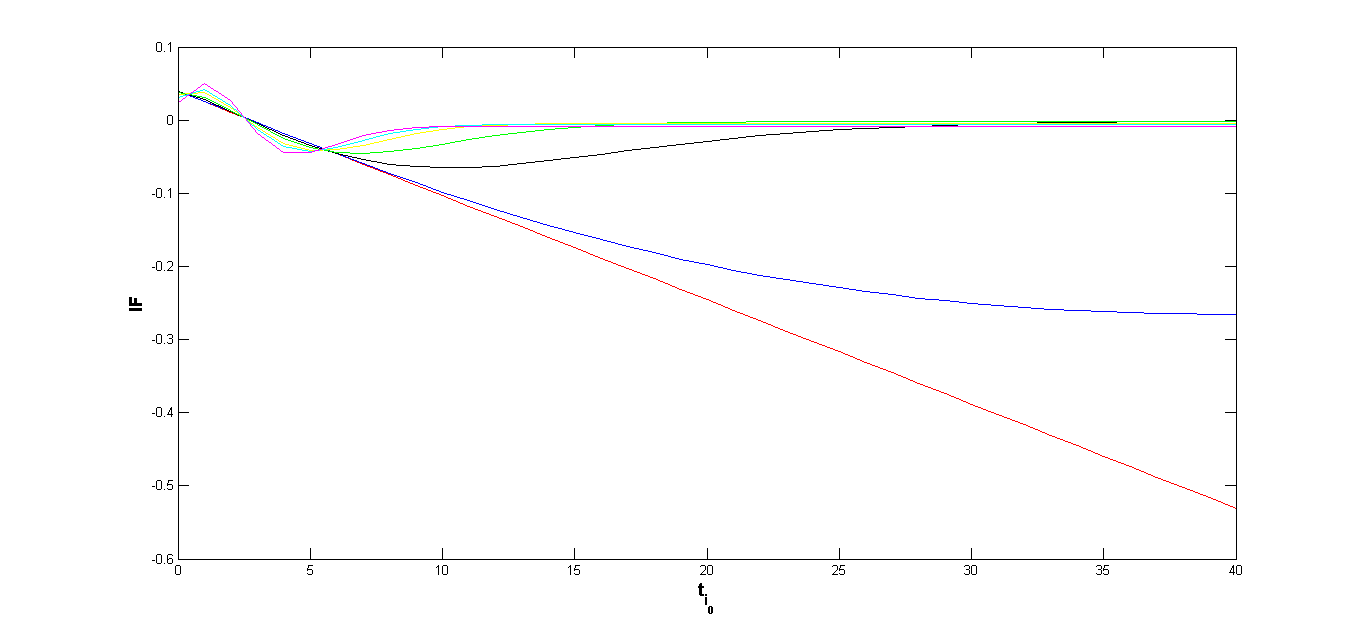}
\label{fig:23_50}}
\\
\subfloat[Model III, $i_0$ = 1]{
\includegraphics[width=0.5\textwidth, height=0.35\textwidth]{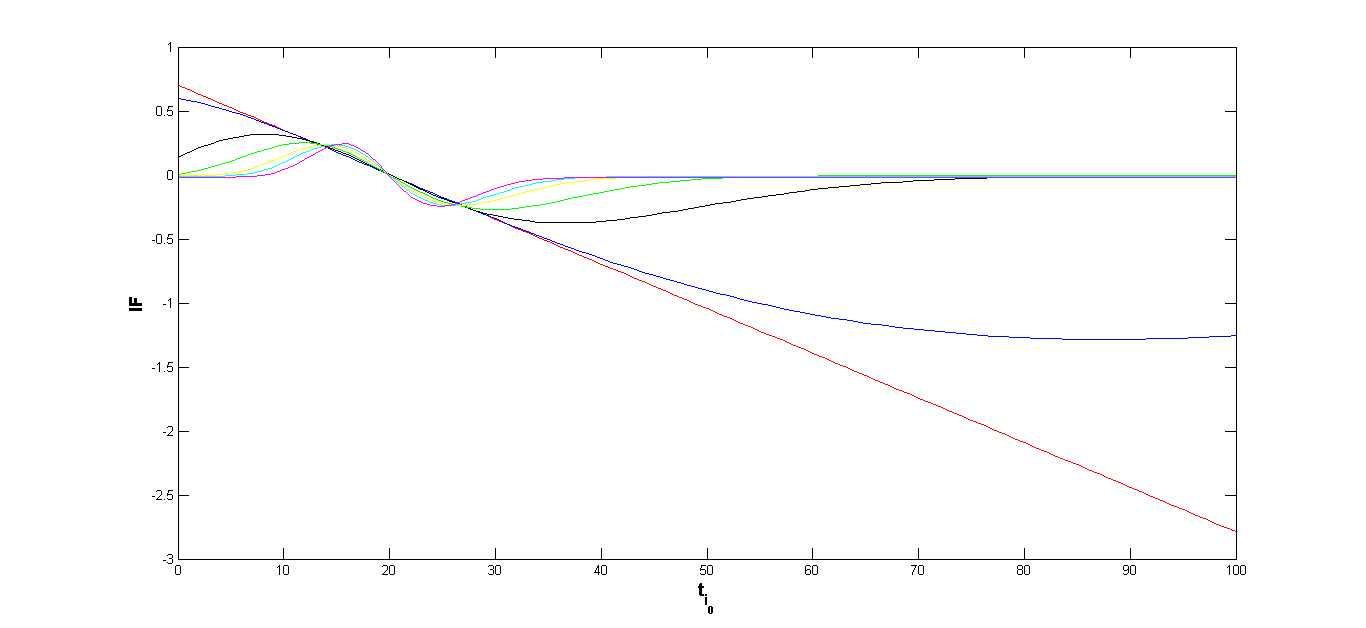}
\label{fig:31_50}}
~ 
\subfloat[Model III, $i_0$ = 50]{
\includegraphics[width=0.5\textwidth, height=0.35\textwidth]{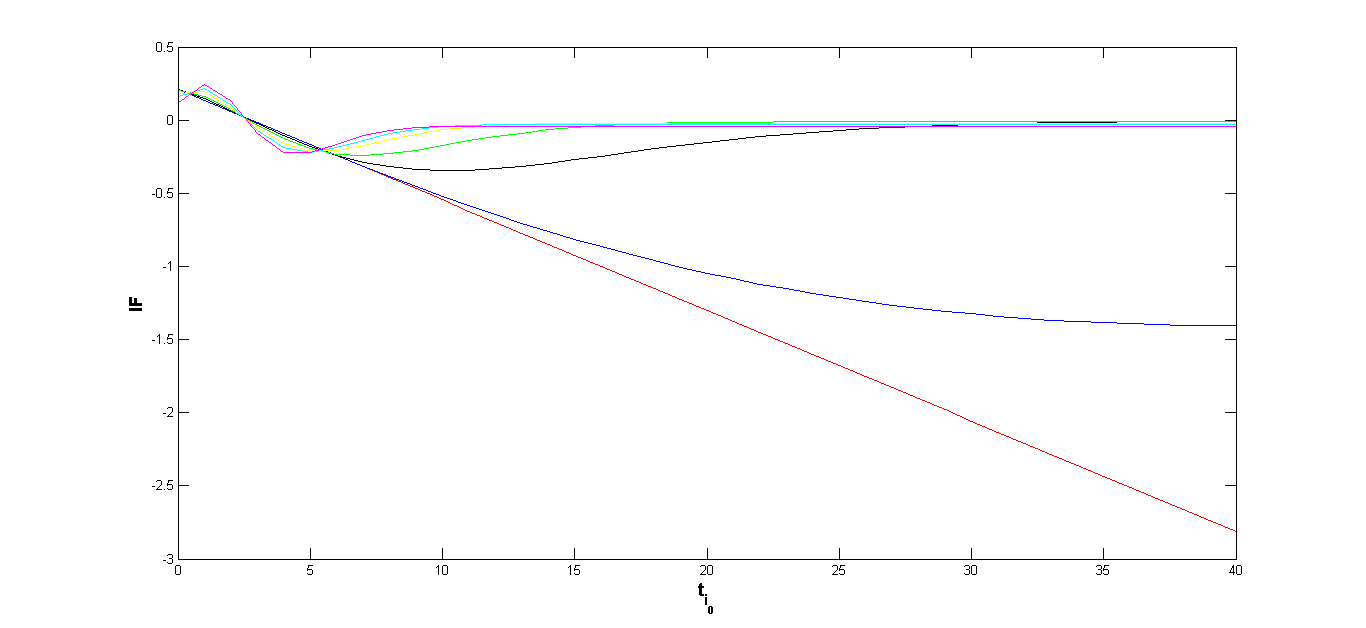}
\label{fig:33_50}}
\caption[Plot of IF of MDPDE of slope parameter $\beta_1$ for different scenario for $n=50$]{Plot of 
the influence function of MDPDE of slope parameter $\beta_1$ for different $\alpha$ and 
direction $i_0$ of contamination in case of three models [Model I : $x_i= (1, \sqrt{i})^T$, 
Model II : $x_i= (1, \frac{1}{i})^T$, Model III : $x_i= (1, \frac{1}{i}, \frac{1}{i})^T$ with 
$\beta_i=1$ and $n=50$]}
\label{FIG:10IF_poisson_50}
\end{figure}

\begin{figure}[h]
\centering
\subfloat[Model I, $i_0$ = 1]{
\includegraphics[width=0.5\textwidth, height=0.35\textwidth]{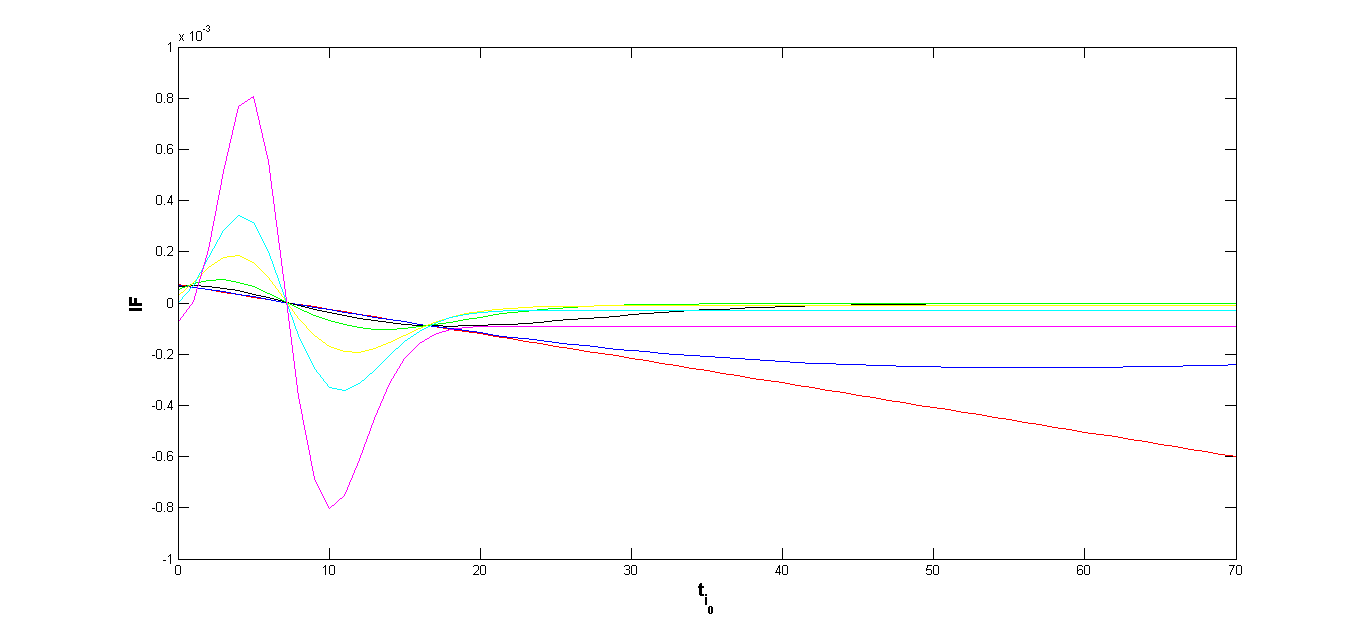}
\label{fig:11_100}}
~ 
\subfloat[Model I, $i_0$ = 50]{
\includegraphics[width=0.5\textwidth, height=0.35\textwidth]{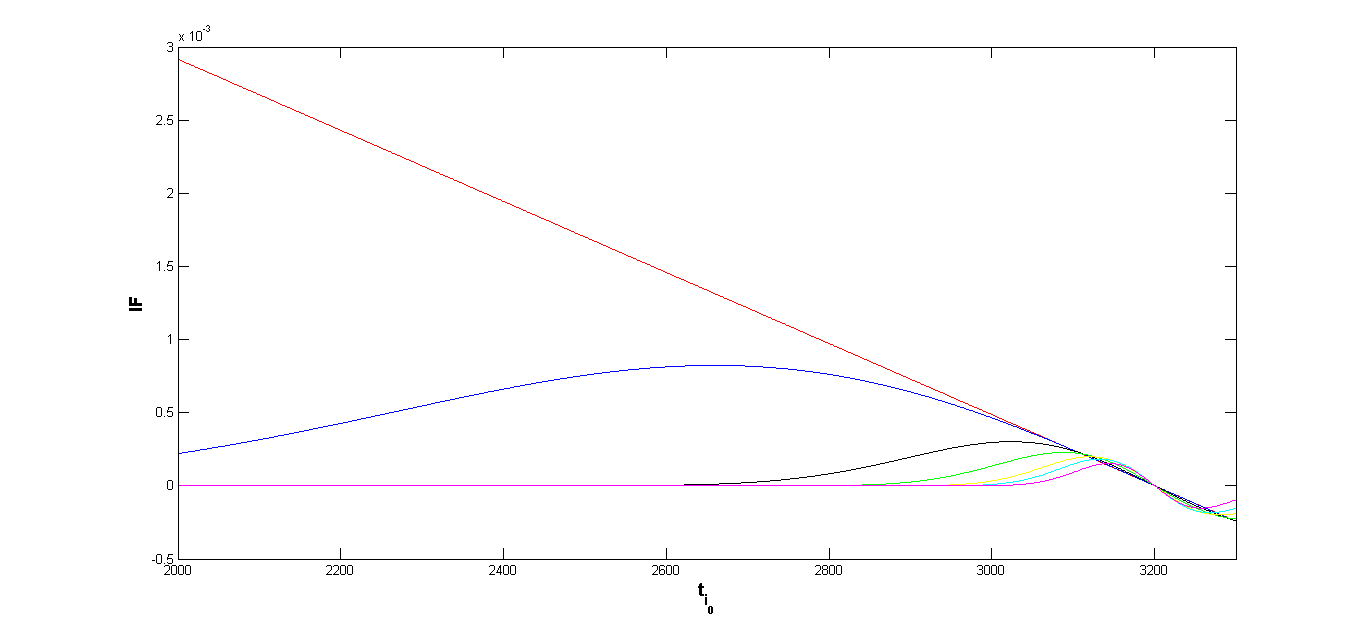}
\label{fig:12_100}}
\\
\subfloat[Model II, $i_0$ = 1]{
\includegraphics[width=0.5\textwidth, height=0.35\textwidth]{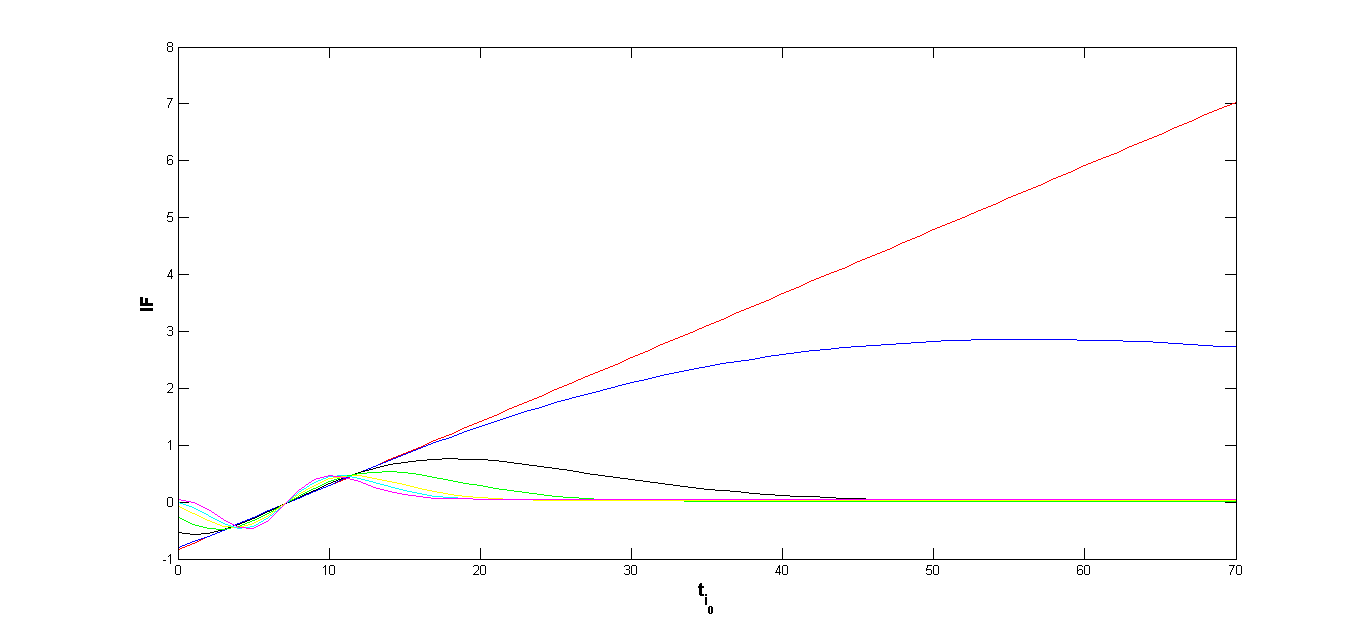}
\label{fig:21_100}}
~ 
\subfloat[Model II, $i_0$ = 50]{
\includegraphics[width=0.5\textwidth, height=0.35\textwidth]{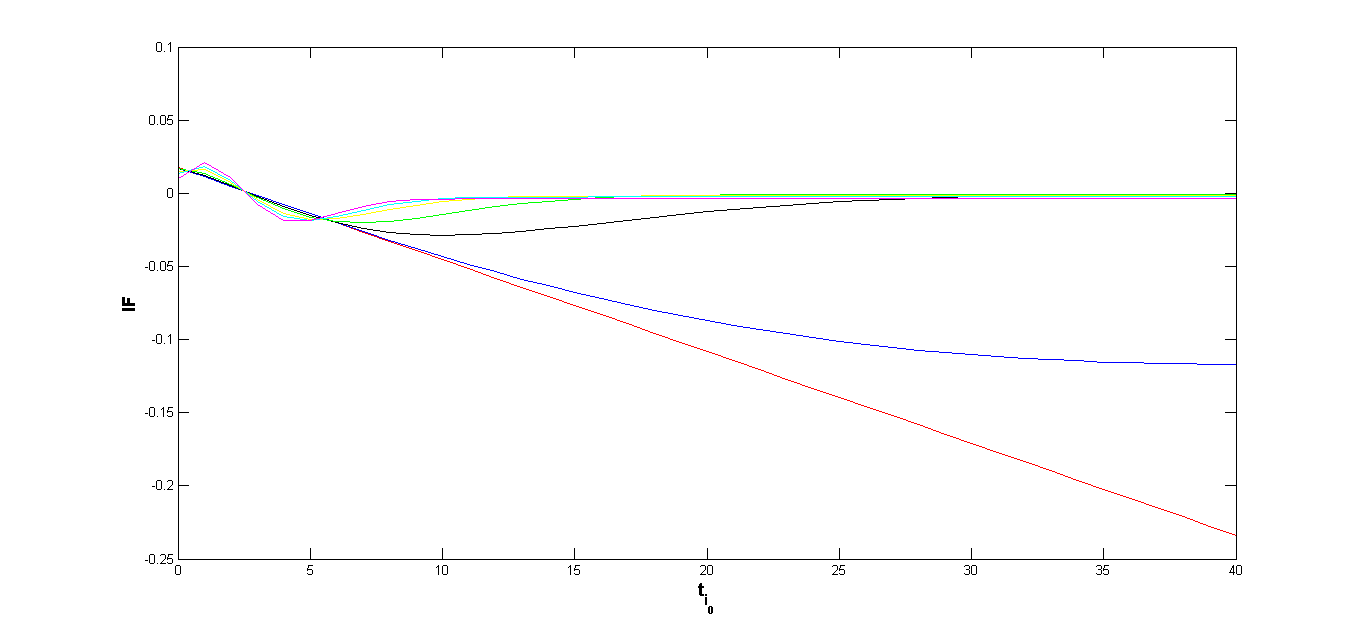}
\label{fig:22_100}}
\\
\subfloat[Model III, $i_0$ = 1]{
\includegraphics[width=0.5\textwidth, height=0.35\textwidth]{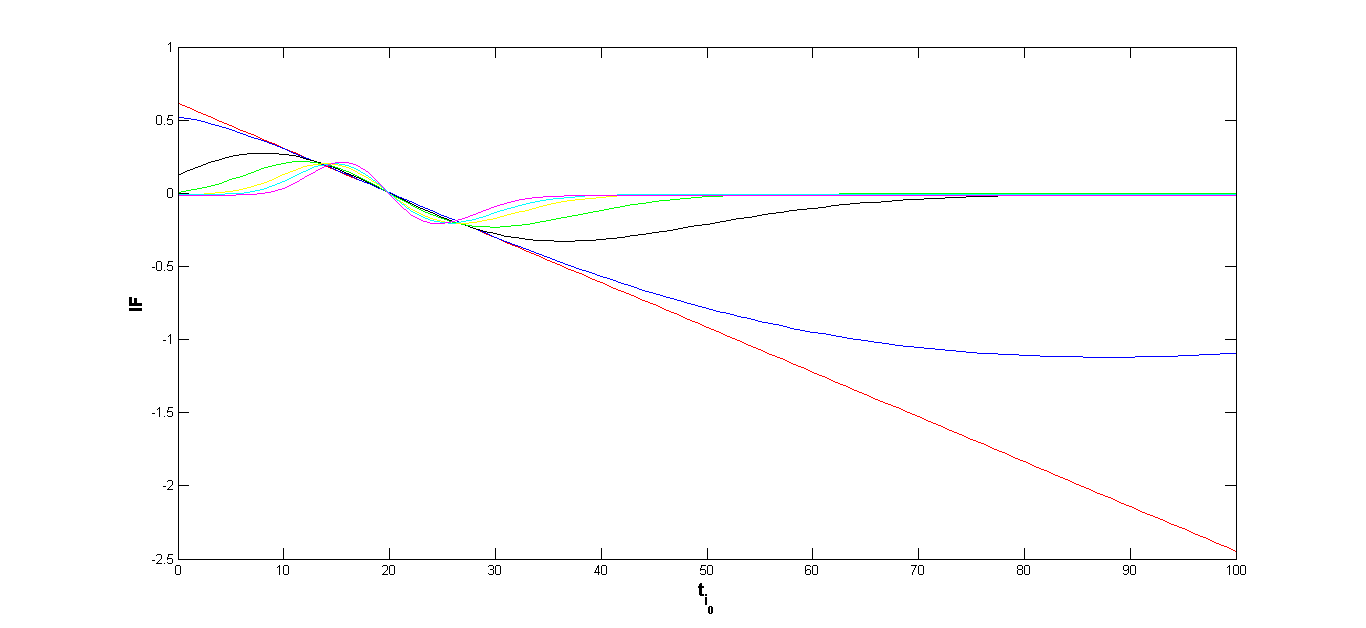}
\label{fig:31_100}}
~ 
\subfloat[Model III, $i_0$ = 50]{
\includegraphics[width=0.5\textwidth, height=0.35\textwidth]{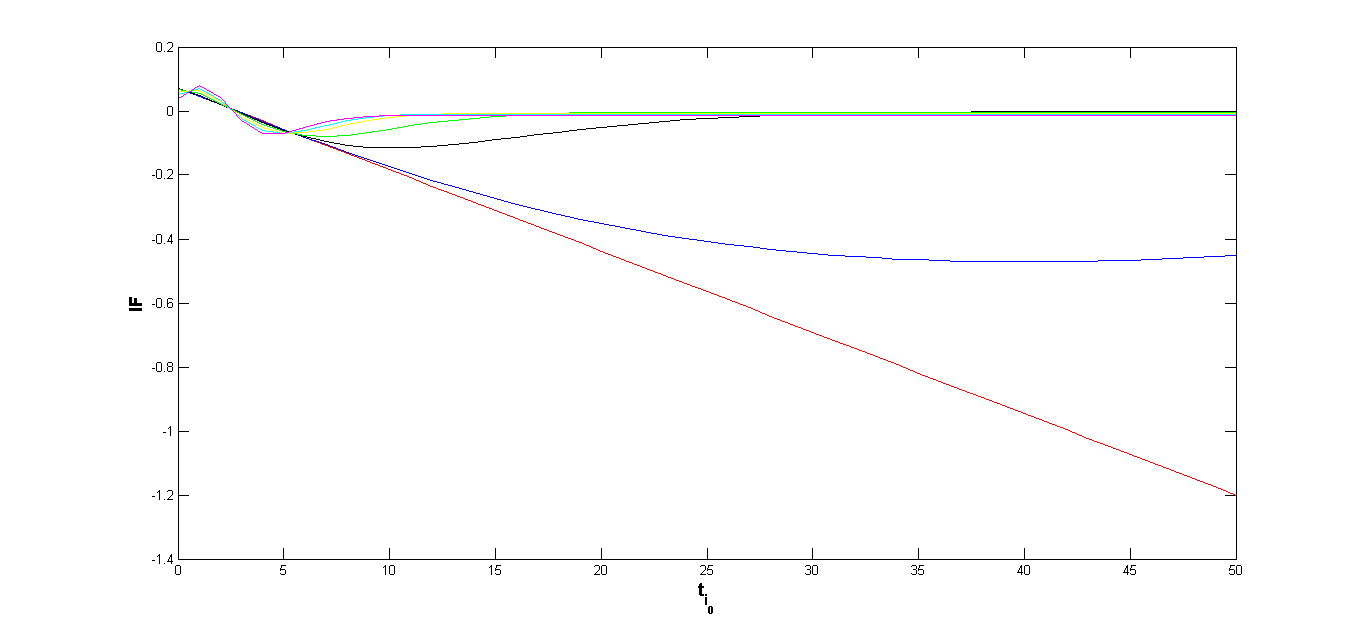}
\label{fig:33_100}}
\caption[Plot of IF of MDPDE of slope parameter $\beta_1$ for different scenario for $n=100$]{Plot of 
the influence functions of MDPDE of slope parameter $\beta_1$ for different $\alpha$ and 
direction $i_0$ of contamination in case of three models [Model I : $x_i= (1, \sqrt{i})^T$, 
Model II : $x_i= (1, \frac{1}{i})^T$, Model III : $x_i= (1, \frac{1}{i}, \frac{1}{i})^T$ with $\beta_i=1$ 
and $n=100$]}
\label{FIG:10IF_poisson_100}
\end{figure}

Clearly, whenever the inverse of the first matrix exists, 
this influence function is bounded in $t_{i_0}$ for any $\alpha > 0$ implying 
the robustness of the minimum density power divergence estimator with $\alpha > 0$. 
However, at $\alpha =0$, the influence function above further simplifies to  
$$
IF_{i_0}(t_{i_0}, T_0^\beta, G_1, \cdots, G_n) =  (X^T\Gamma_{11}^{(0)}X)^{-1} x_{i_0} 
\left(t_{i_0} - e^{(x_{i_0}^T\beta)}\right) ,
$$
which is linear and hence unbounded in $t_{i_0}$. This indicates the non-robustness of the maximum likelihood estimator 
and equivalently ordinary least squares of the regression parameter in case of the Poisson regression model. 
Figures \ref{FIG:10IF_poisson_50} and \ref{FIG:10IF_poisson_100} show the Influence function of 
the minimum density power divergence estimator for different $\alpha$ under several specific Poisson 
regression models and for sample sizes $50$ and $100$ respectively. The redescending nature of the influence 
function with increasing $\alpha$ is quite clear in all the figures.

\newpage
\clearpage
\newpage

\clearpage

\section{Special Case (II) : Logistic Regression for Binary Data}\label{SEC:MDPDE_logistic}

Another important special case of the generalised linear model is the logistic regression model which is used
to model any categorical or binary dependent variables in terms of some explanatory variable. 
Given the values of the explanatory variables $x_i$, the binary outcome variable $y_i$ 
(or the binary transform of the categorical variables) are assumed to follow a Bernoulli trial 
with success probability $\pi_i$ depending on the  explanatory variable $x_i$ (for each $i=1, \cdots, n$). 
To ensure that the predicted values of $\pi_i$ are  in 
 the interval $(0, 1)$, in the logistic model it is assumed that 
 
 $$
 \pi_i = \pi(x_i) = \frac{e^{x_i^T\beta}}{1 + e^{x_i^T\beta}}.
 $$ 
 We will now assume that the $x_i$'s are fixed and consider the logistic regression model from its
 design perspective to estimate $\beta$ efficiently and robustly.

\subsection{The minimum density power divergence estimator for the Logistic Regression}

	We can treat the logistic regression model as a particular case of the generalised linear model with 
known shape parameter $\phi=1$ and $\theta_i=\eta_i=x_i^T\beta$, $c(y_i)=0$. The distribution of $y_i$ is 	
the Bernoulli distribution with mean 
	$\mu_i = \pi_i =  \frac{e^{\eta_i}}{1 + e^{\eta_i}},$
and $var(y_i) = \pi_i(1-\pi_i) = \frac{e^{\eta_i}}{(1 + e^{\eta_i})^2}.$
Thus the link function $g$ is the logit function and so we can use the minimum density power divergence 
estimation procedure discussed in Section \ref{SEC:MDPDE_GLM} to estimate $\beta$ robustly.
Using the above notations and the form of the Bernoulli distribution, a simple calculation yields
	$K_{1i}(y_i; \beta) = (y_i - \mu_i)$ so that 
\begin{eqnarray}
\gamma_{1i} &=& (1-\mu_i)\mu_i^{1+\alpha} - \mu_i (1-\mu_i)^{1+\alpha} 
	= \frac{e^{x_i^T\beta} (e^{\alpha(x_i^T\beta)} - 1)}{(1 + e^{x_i^T\beta})^{2+\alpha}},\nonumber\\
\mbox{and }~~~~~~~~ \gamma_{11i} &=& (1-\mu_i)^2\mu_i^{1+\alpha} + \mu_i^2 (1-\mu_i)^{1+\alpha} 
	= \frac{e^{x_i^T\beta} (e^{\alpha(x_i^T\beta)} + e^{x_i^T\beta})}{(1 + e^{x_i^T\beta})^{3+\alpha}}.
\end{eqnarray}
Then the minimum density power divergence estimating equation for $\alpha \geq 0$ is given by,
\begin{equation}
\sum_{i=1}^n \left[  \frac{e^{x_i^T\beta} (e^{\alpha(x_i^T\beta)} - 1)}{(1 + e^{x_i^T\beta})^{2+\alpha}} 
- \left(y_i -  \frac{e^{x_i^T\beta} }{1 + e^{x_i^T\beta}}\right)  \frac{e^{\alpha(x_i^T\beta)y_i} }{(1 + 
e^{x_i^T\beta})^{\alpha}}\right]x_i = 0, \label{EQ:Est_eqn_Logistic1}
\end{equation}
which can be further simplified to 
\begin{equation}
\sum_{i=1}^n (1-2y_i)e^{(x_i^T\beta)(1-y_i)} ~ \frac{ (e^{\alpha(x_i^T\beta)} + e^{x_i^T\beta})}{(1 + 
e^{x_i^T\beta})^{2+\alpha}} x_i = 0. \label{EQ:Est_eqn_Logistic2}
\end{equation}
We can easily solve the above estimating equation with respect to $\beta$ to compute the minimum density power divergence estimator for any 
$\alpha \geq 0$. In particular, for $\alpha=0$, equation (\ref{EQ:Est_eqn_Logistic1}) simplifies to 
\begin{equation}
\sum_{i=1}^n \left(y_i -  \frac{e^{x_i^T\beta} }{1 + e^{x_i^T\beta}}\right) x_i = 0, 
\label{EQ:Est_eqn_Logistic0}
\end{equation}
which is the maximum likelihood estimating equation. Once again the minimum density power divergence estimator estimating equation is just a 
generalization of the maximum likelihood estimating equation.

\subsection{Properties of minimum density power divergence estimator }

We will now present the asymptotic distribution of the minimum density power divergence estimator of $\beta$ in the logistic regression case
as it follows from Theorem  \ref{THM:asymp_GLM}. In this special case, we have
$$
\Psi_n(\beta) 
=\frac{1}{n}\sum_{i=1}^n~e^{x_i^T\beta}\frac{(e^{\alpha(x_i^T\beta)}+e^{x_i^T\beta})}{(1+e^{x_i^T\beta})^{3+\alpha}}
(x_i x_i^T)
$$
$$
\mbox{and }~~~~~ \Omega_n(\beta) = \frac{1}{n} \sum_{i=1}^n ~ e^{x_i^T\beta} \frac{ (e^{\alpha(x_i^T\beta)} 
+ e^{x_i^T\beta})^2}{(1 + e^{x_i^T\beta})^{4+2\alpha}} (x_i x_i^T).
$$
We then have the following result.
\begin{corollary}
Under Assumptions (A1)-(A7) of Ghosh and Basu (2013a), there exists a consistent sequence 
$\hat\beta_n = \hat\beta_n^{(\alpha)}$  of roots to the minimum density power divergence estimating equations 
(\ref{EQ:Est_eqn_Logistic2}) at the tuning parameter $\alpha$. Also, the asymptotic distribution of
\begin{eqnarray}
&& ~~ \left(\sum_{i=1}^n ~ e^{x_i^T\beta^g} \frac{ (e^{\alpha(x_i^T\beta^g)} + e^{x_i^T\beta^g})^2}{(1 + 
e^{x_i^T\beta^g})^{4+2\alpha}} (x_i x_i^T)\right)^{-\frac{1}{2}} \times \nonumber \\
&& ~~~~~~~~~~~~~~~~~~~~~~~~~~~~~~~~~~~~~~~~~\left(\sum_{i=1}^n ~ e^{x_i^T\beta^g} \frac{(e^{\alpha(x_i^T\beta^g)} 
+ e^{x_i^T\beta^g})}{(1 + e^{x_i^T\beta^g})^{3+\alpha}} (x_i x_i^T)\right)(\hat\beta_n - \beta^g) \nonumber
\end{eqnarray}
is	 $p$-dimensional normal with mean $0$ and variance $I_{p}$.
\end{corollary}
	
Then, as argued in Section \ref{SEC:asymp_poisson} for the Poisson regression, the asymptotic efficiency of 
the different minimum density power divergence estimator $\hat\beta_n=\hat\beta_n^{(\alpha)}$ of $\beta$ for the logistic regression can also 
be measured in terms of its asymptotic variance 
	\begin{eqnarray}
	&& AV_\alpha(\beta^g) \nonumber \\
	&&~~= \left(\sum_{i=1}^n ~ e^{x_i^T\beta^g} \frac{ (e^{\alpha(x_i^T\beta^g)} + e^{x_i^T\beta^g})}{(1 + e^{x_i^T\beta^g})^{3+\alpha}} (x_i x_i^T)\right)^{-1} \left(\sum_{i=1}^n ~ e^{x_i^T\beta^g} \frac{ (e^{\alpha(x_i^T\beta^g)} + e^{x_i^T\beta^g})^2}{(1 + e^{x_i^T\beta^g})^{4+2\alpha}} (x_i x_i^T)\right) \nonumber \\
	 & & ~~~~~~~~~~~~~~~~~~~~~~~~~~~~~~~ \left(\sum_{i=1}^n ~ e^{x_i^T\beta^g} \frac{ (e^{\alpha(x_i^T\beta^g)} + e^{x_i^T\beta^g})}{(1 + e^{x_i^T\beta^g})^{3+\alpha}} (x_i x_i^T)\right)^{-1}.
	\end{eqnarray}
	This can be estimated consistently by $\widehat{AV}_\alpha = AV_\alpha(\hat\beta_n)$.

	As in the Poisson regression case, here also we can compute the values of relative efficiencies of the 
minimum density power divergence estimators of the coefficients of the logistic regression model based on $\widehat{AV}_\alpha$. 
This measure of relative efficiency clearly depends on the value of $\beta$ and $X_i$s. 
We present the empirical estimate of the relative efficiencies of the MDPDE in case of the 
logistic regression model in Tables \ref{TAB:ARE_poiss_50} and \ref{TAB:ARE_poiss_100} respectively 
 for sample size $n=50$ and $n=100$. These are calculated based on a simulation study based on 
 $1000$ replications under several different cases of logistic regressions. 
 These cases are defined based on the true values of the regression coefficients 
 $\beta=(\beta_0, \beta_1, \ldots, \beta_p)$ and 
the given values of the explanatory variables $x_i$ ($i=1,\ldots,n$) as follows:
\begin{eqnarray}
\mbox{Case I~~} &:& p=2; \beta=(0.1, 0.1) \mbox{  and  } x_i = (1, \sqrt{i}). \nonumber\\
\mbox{Case II~} &:& p=2; \beta=(0.001, 0.0001) \mbox{ and } x_i = (1, \sqrt{i}). \nonumber\\
\mbox{Case III} &:&  p=2; \beta=(1, 1) \mbox{  and  } x_i = (1, \frac{1}{i}). \nonumber\\
\mbox{Case IV} &:&  p=2; \beta=(0.1, 0.1) \mbox{  and  } x_i = (1, \frac{1}{i}). \nonumber
\end{eqnarray}
\begin{eqnarray}
\mbox{Case V~} &:&  p=3; \beta=(0.1, 0.1, 0.1) \mbox{  and  } x_i = (1, \sqrt{i}, \frac{1}{i^2}).\nonumber\\
\mbox{Case VI} &:&  p=3; \beta=(0.01, 0.001, 0.0001) \mbox{  and  } x_i = (1, \sqrt{i}, \frac{1}{i^2}).\nonumber
\end{eqnarray}
It is clearly seen from the tables that for any value 
of the parameter and the explanatory variables, the loss of efficiency is negligible for small $\alpha>0$. 
Further, if the values of $x_i^T\beta$ is small, then we can get very quite high efficiency even for large
positive $\alpha$ near $0.5$.

\begin{table}[h]
\caption{The estimated Relative efficiencies of the MDPDE for various values of the tuning parameter $\alpha$  under different cases of Logistic Regression with sample size $n = 50$}
\label{TAB:ARE_logistic_50}
\begin{center}
\resizebox{0.8\textwidth}{!}{	
\begin{tabular}{llrrrrrrrr} \hline
Case	& Coefficients	&	$\alpha = 0$	&	$\alpha = 0.01$	&	$\alpha = 0.1$	&	$\alpha = 0.25$	&	$\alpha = 0.4$	&	$\alpha = 0.5$	&	$\alpha = 0.7$	&	$\alpha = 1$
	       			\\	 \hline	       		
I	&	$\beta_0$	&	100.0	&	99.0	&	90.7	&	74.6	&	67.6	&	61.3	&	50.4	&	37.5	\\	 
	&	$\beta_1$	&	100.0	&	99.2	&	92.7	&	79.6	&	73.8	&	68.4	&	58.7	&	46.7	\\	 \hline
II	&	$\beta_0$	&	100.0	&	99.3	&	93.3	&	81.2	&	75.8	&	70.7	&	61.5	&	50.0	\\	 
	&	$\beta_1$	&	100.0	&	99.3	&	93.3	&	81.2	&	75.8	&	70.7	&	61.5	&	50.0	\\	 \hline
III	&	$\beta_0$	&	100.0	&	98.6	&	86.7	&	65.2	&	56.5	&	49.0	&	36.8	&	23.9	\\	 
	&	$\beta_1$	&	100.0	&	98.1	&	82.8	&	56.9	&	47.2	&	39.2	&	27.1	&	15.6	\\	 \hline
IV	&	$\beta_0$	&	100.0	&	99.3	&	92.8	&	79.8	&	74.0	&	68.7	&	59.1	&	47.2	\\	 
	&	$\beta_1$	&	100.0	&	99.2	&	92.4	&	79.0	&	73.0	&	67.5	&	57.7	&	45.6	\\	 \hline
V	&	$\beta_0$	&	100.0	&	99.3	&	92.7	&	79.8	&	74.0	&	68.6	&	59.0	&	47.1	\\	 
	&	$\beta_1$	&	100.0	&	99.2	&	92.6	&	79.4	&	73.5	&	68.1	&	58.4	&	46.3	\\	 \hline
	&	$\beta_2$	&	100.0	&	99.2	&	92.4	&	78.9	&	72.9	&	67.4	&	57.5	&	45.4	\\	 
VI	&	$\beta_0$	&	100.0	&	99.3	&	93.3	&	81.1	&	75.6	&	70.5	&	61.3	&	49.7	\\	 \hline
	&	$\beta_1$	&	100.0	&	99.3	&	93.3	&	81.1	&	75.6	&	70.5	&	61.3	&	49.7	\\	 
	&	$\beta_2$	&	100.0	&	99.3	&	93.3	&	81.1	&	75.6	&	70.5	&	61.3	&	49.7		\\	\hline
\end{tabular}}
\end{center}
\end{table}

\begin{table}[h] 
\caption{The estimated Relative efficiencies of the MDPDE for various values of the tuning parameter $\alpha$  under different cases of Logistic Regression with sample size $n = 50$}
\label{TAB:ARE_logistic_100}
\begin{center}
\resizebox{0.8\textwidth}{!}{
\begin{tabular}{llrrrrrrrr} \hline
Case	& Coefficients	&	$\alpha = 0$	&	$\alpha = 0.01$	&	$\alpha = 0.1$	&	$\alpha = 0.25$	&	$\alpha = 0.4$	&	$\alpha = 0.5$	&	$\alpha = 0.7$	&	$\alpha = 1$
	\\	 \hline
I	&	$\beta_0$	&	100.0	&	98.9	&	89.8	&	72.3	&	64.9	&	58.2	&	46.8	&	33.7	\\	 
	&	$\beta_1$	&	100.0	&	99.2	&	92.6	&	79.5	&	73.6	&	68.2	&	58.5	&	46.5	\\	 \hline
II	&	$\beta_0$	&	100.0	&	99.3	&	93.3	&	81.2	&	75.8	&	70.7	&	61.5	&	50.0	\\	 
	&	$\beta_1$	&	100.0	&	99.3	&	93.3	&	81.2	&	75.8	&	70.7	&	61.5	&	50.0	\\	 \hline
III	&	$\beta_0$	&	100.0	&	98.6	&	87.0	&	65.9	&	57.4	&	49.9	&	37.8	&	24.8	\\	 
	&	$\beta_1$	&	100.0	&	98.1	&	82.9	&	57.2	&	47.5	&	39.6	&	27.4	&	15.9	\\	 \hline
IV	&	$\beta_0$	&	100.0	&	99.3	&	92.8	&	79.9	&	74.1	&	68.8	&	59.2	&	47.3	\\	 
	&	$\beta_1$	&	100.0	&	99.2	&	92.4	&	79.0	&	73.0	&	67.5	&	57.7	&	45.6	\\	 \hline
V	&	$\beta_0$	&	100.0	&	99.3	&	92.8	&	79.8	&	74.1	&	68.7	&	59.1	&	47.2	\\	 
	&	$\beta_1$	&	100.0	&	99.2	&	92.6	&	79.4	&	73.6	&	68.1	&	58.4	&	46.4	\\	 \hline
	&	$\beta_2$	&	100.0	&	99.2	&	92.4	&	78.9	&	72.9	&	67.3	&	57.5	&	45.3	\\	 
VI	&	$\beta_0$	&	100.0	&	99.3	&	93.3	&	81.1	&	75.6	&	70.5	&	61.3	&	49.7	\\	 \hline
	&	$\beta_1$	&	100.0	&	99.3	&	93.3	&	81.1	&	75.6	&	70.5	&	61.3	&	49.7	\\	 
	&	$\beta_2$	&	100.0	&	99.3	&	93.3	&	81.1	&	75.6	&	70.5	&	61.3	&	49.7	\\	\hline
\end{tabular}}
\end{center}
\end{table}

\section{A Data-driven Choice of the tuning parameter $\alpha$}\label{SEC:alpha_GLM}

The minimum density power divergence estimators depend on the choice of the tuning parameter $\alpha \geq 0$
defining the divergence.  The properties of the minimum density power divergence estimator in the case of independent and identically 
distributed data have been extensively studied in the literature and it is well known that 
there is a trade off between efficiency and robustness for varying $\alpha$. 
Increasing $\alpha$ leads to greater robustness at the cost of efficiency.
Ghosh and Basu (2013a, 2013b) also observed similar trade-offs for the linear regression case with 
fixed covariates. In the two previous sections, we have observed the same phenomenon in the context of
 the proposed minimum density power divergence estimator for the Poisson and the logistic regression models. Therefore, it is necessary to  
 carefully choose the tuning parameter $\alpha$ while using the minimum density power divergence estimator in any of the generalised 
 linear regression models. In this section, we will try to present a possible approach to choose the 
 optimum value of $\alpha$  based on the observed data at hand.

In the context of the i.i.d. data problems, some data driven choices for selecting the optimum tuning 
parameter in the minimum density power divergence estimation context have been proposed 
by Hong and Kim (2001) and Warwick and Jones (2005). Ghosh and Basu (2013b) extended these 
approaches to the case of independent but non-homogeneous data and illustrated this approach 
for the case of linear regression through detailed simulation studies. In this present paper, 
we consider the generalised linear model from its design perspective so that given the values 
of the explanatory variables $x_i$s the response $y_i$'s are independent but not identically distributed. 
So, we can apply the results of Ghosh and Basu (2013b) to 
choose a data-driven optimum choice of the tuning parameter $\alpha$. Accordingly, we need to 
choose $\alpha$ by minimizing a consistent  estimate of the mean square error (MSE) of the minimum density power divergence estimator
$\hat{\theta} = (\hat{\beta}_\alpha,~ \hat{\phi}_\alpha)$ of the true parameter value 
$\theta^g=(\beta^g,~\phi^g)$, defined as  
$E\left[ (\hat{\theta}_\alpha - \theta^*)^T(\hat{\theta}_\alpha - \theta^*)\right]$,  
in generalised linear model. But, it follows from the asymptotic distribution of the minimum density power divergence estimator that, 
asymptotically
\begin{equation}
E\left[ (\hat{\theta}_\alpha - \theta^g)^T(\hat{\theta}_\alpha - \theta^g)\right] 
= ({\theta}_\alpha - \theta^g)^T({\theta}_\alpha - \theta^g) 
+ \frac{1}{n} Trace\left[\Psi_n^{-1} \Omega_n \Psi_n^{-1} \right],
\label{EQ:MSE_tuning_parameter_nonhomogeneous}
\end{equation}
where $\Psi_n$ and $\Omega_n$ are as defined in Section \ref{SEC:asymp_GLM} and 
$\theta_\alpha = (\beta_\alpha,~\phi_\alpha)$ is the parameter value minimizing the 
density power divergence corresponding to parameter $\alpha$. Further, from the 
expressions of $\Psi_n$ and $\Omega_n$  it is sufficient to find some consistent estimator of 
the quantities $\gamma_{ji}$ and $\gamma_{jki}$ for $j, k = 1,2$ and $ i=1, \cdots, n$, 
which can be done by replacing the parameter value $(\beta,~\phi)$ in their expressions by 
the corresponding minimum density power divergence estimators 
$(\hat{\beta}_\alpha,~ \hat{\phi}_\alpha)$. Let us denote the resulting matrices by $\hat{\Psi}_n$
 and $\hat{\Omega}_n$. To estimate the bias term, we will use $(\hat{\beta}_\alpha,~ \hat{\phi}_\alpha)$ 
 as a consistent estimate of the $({\beta}_\alpha,~{\phi}_\alpha )$. 
 For estimating $\theta^*$, we can use several ``pilot"  estimators which will in turn affect the  
 final choice of the tuning parameter. Ghosh and Basu (2013b) suggested, on the basis of 
 an extensive simulation study that the choice of minimum density power divergence estimator with $\alpha =0.5$ as a reasonable choice. 
 For any particular generalised model reasonable pilot estimators may be obtained by simulation.

In the two special cases of generalised linear model discussed above, namely the Poisson and the 
logistic regression models, the parameter $\phi$ is known so that $\theta=\beta$ and the expressions of
 $\Psi_n$ and $\Omega_n$ are also explicitly derived in the respective sections. So, we can 
 apply the above discussed method to get a data-driven optimum value of the tuning parameter $\alpha$
 for these two models. 
 For any particular sample data, we can apply the above method with different ``pilot" estimates 
 and then choose one having the minimum estimated MSE value.
 It is observed from numerical studies that it is also sufficient to consider
 the minimum density power divergence estimator corresponding to $\alpha=0.5$ as ``pilot" estimator 
 in case of Poisson and logistic regression models also.

For simplicity, we will present only the results corresponding to the particular real data examples 
considered in this paper.
Table \ref{TAB:opt_alpha} at the end of this paper provides the optimum value of $\alpha$ 
obtained by minimising the estimated MSE based on several ``Pilot" $\alpha$ for all these data sets. 
We have already seen that there are some values of the tunning parameter $\alpha$ 
for which the MDPDE of the parameters become robust for the respective data set. 
Now, comparing it with the table of optimum $\alpha$, 
we see that the choice of pilot $\alpha$ is very crucial in order to get a optimum $\alpha$ 
leading to a robust estimator.  In this regard, the choice $\alpha=0.5$ as the pilot $\alpha$ 
is sufficient to give us the ``good" optimum $\alpha$ values for each of the data set and leads to the 
robust estimators of the parameter in next stage. Thus, we suggest $\alpha=0.5$ as a reasonable choice 
for ``pilot" $\alpha$ for obtaining the MDPDE in generalized linear regression.
Note that, this choice is also in-line with the work of Ghosh and Basu (2013b) where the same choice 
for the pilot $\alpha$ was proposed in the context of linear regression model.

\section{Real Data Examples}\label{SEC:Data_examples}

In this section, we will explore the performance of the proposed minimum density power divergence estimators in 
Poisson and logistic regression models by applying it on some interesting real data-sets. 
These data-sets are obtained from several clinical trials or surveys which are very important 
in biological sciences to demonstrate the importance of the proposed methodology 
in that field of application.

\subsection{Epilepsy Data}

In this example we will consider an interesting data-set consisting of  59 epilepsy patients from 
Thall and Vail (1990). The data  were obtained from a clinical trial carried out by Leppik et al.~(1985)
 where the patients were treated by the anti-epileptic drug ``prog-abide" or a placebo with 
 randomized assignment. Then the total number of epilepsy attacks was noted which can be modeled by the Poisson 
 distribution. Here, we will try to explain the total number of attacks by an appropriate set of 
 explanatory variables through a Poisson regression model (Hosseinian, 2009). 
 The variables considered in this regard are ``Base", 
 the eight week baseline seizure rate prior to randomization in multiples of 4, ``Age", 
 patients' age in multiple of 10 years, and ``Trt", an binary indicator for the treatment-control group.
 Also, the interaction between treatment and baseline seizure rate is important in this case, 
 because it represents either higher or lower seizure rate for the treatment group compared to the placebo
 group depending on the baseline count. In fact, the drug decreases the epilepsy only if the baseline count becomes
 sufficiently large in numbers with respect to some critical threshold.

   	The data were also analyzed by Hosseinian (2009) who compared the maximum likelihood estimator with the robust methodologies
proposed by herself in the same paper and those by Cantoni and Ronchetti (2001). There it was observed that 
the data contain some outlying observations for which the interaction effect between treatment and 
baseline seizure rate turns out to be insignificant based on  the maximum likelihood estimator whereas the robust estimators 
   	show this interaction to be significant. Here, we will apply our proposed robust minimum density power divergence estimators 
   	for this epilepsy data set and try to see if our proposed estimators are also robust enough to 
   	differentiate with maximum likelihood estimator for the interaction effect.

  	Table \ref{TAB:est_epilepsy_dat} presents the parameter estimates,
  	their asymptotic standard errors and corresponding p-values based on the minimum density power divergence estimator with different $\alpha$.
  	Clearly the estimators corresponding to $\alpha \geq 0.3$ are quite different from maximum likelihood estimator and 
  	for these estimators the interaction effect is also significant under the Poisson regression model. 
  	Indeed, these estimators are quite similar to the robust estimators considered in Hosseinian (2009) 
  	but have greater asymptotic efficiency.

\begin{table}[h]
\caption{The minimum density power divergence estimates, their standard errors and p-values for the Epilepsy Data }
\label{TAB:est_epilepsy_dat}
\begin{center}
\begin{tabular}{llrrrrrr} \hline
		&		&	$\alpha=0$	&	$\alpha=0.1$	&	$\alpha=0.3$	&	$\alpha=0.5$	&	$\alpha=0.7$	&	$\alpha=1$	\\ \hline
	Intercept	&	Estimate	&	1.9888	&	2.1089	&	1.9106	&	1.9691	&	2.0060	&	1.9653	\\
		&	SE ($\times 100$)	&	13.6518	&	15.2509	&	12.6869	&	13.7081	&	14.9185	&	17.0043	\\
		&	P-Value	&	0.0000	&	0.0000	&	0.0000	&	0.0000	&	0.0000	&	0.0000	\\ \hline
	Trt	&	Estimate	&	-0.2375	&	-0.3169	&	-0.3871	&	-0.3893	&	-0.3516	&	-0.3186	\\
		&	SE ($\times 100$)	&	7.6816	&	8.6812	&	7.9139	&	8.4566	&	9.1111	&	10.1787	\\
		&	P-Value	&	0.0030	&	0.0006	&	0.0000	&	0.0000	&	0.0003	&	0.0027	\\ \hline
	Base	&	Estimate	&	0.0858	&	0.0866	&	0.1689	&	0.1631	&	0.1622	&	0.1562	\\
		&	SE ($\times 100$)	&	0.3698	&	0.4101	&	0.2778	&	0.3055	&	0.3359	&	0.3959	\\
		&	P-Value	&	0.0000	&	0.0000	&	0.0000	&	0.0000	&	0.0000	&	0.0000	\\ \hline
	Age	&	Estimate	&	0.2308	&	0.1153	&	0.0408	&	0.0362	&	0.0242	&	0.0559	\\
		&	SE ($\times 100$)	&	4.1498	&	4.7242	&	3.9374	&	4.2416	&	4.6138	&	5.2119	\\
		&	P-Value	&	0.0000	&	0.0177	&	0.3045	&	0.3972	&	0.6017	&	0.2878	\\ \hline
Trt$\times$Base	&	Estimate	&	0.0069	&	0.0107	&	0.0156	&	0.0165	&	0.0131	&	0.0098	\\
		&	SE ($\times 100$)	&	0.4443	&	0.4893	&	0.3230	&	0.3537	&	0.3888	&	0.4600	\\
		&	P-Value	&	0.1283	&	0.0323	&	0.0000	&	0.0000	&	0.0013	&	0.0373	\\\hline
\end{tabular}\vspace{3pt}
\end{center} 
\end{table}

\subsection{Australia AIDS Data}

We will consider an interesting data set on the number of AIDS patients in Australia 
by the date of diagnosis for successive quarters of 1984 to 1988 (Dobson, 2002). This is a part of 
a large survey by National Centre for HIV Epidemiology and Clinical Research and reported in 1994.
Dobson (2002) modeled these count data using a Poisson regression model with the logarithm of time as
the explanatory variable. 
Indeed, there is no outlier in the data. We will first apply the proposed MDPDE for 
different $\alpha$ to illustrate its performance in the absence of outliers. 
The estimates obtained are given in Table \ref{TAB:est_AIDS_Data} and 
are very close to the MLE (corresponding to $\alpha=0$) although their standard error increases slightly with increasing $\alpha>0$.

 \begin{table}[!t]
\caption{The minimum DPD estimates of the parameters for the Australia AIDS Data. The standard error of the estimates are given in the parenthesis.}
\label{TAB:est_AIDS_Data}
\begin{center}
\begin{tabular}{llrrrrrr}\hline
	&	$\alpha$	&	0	&	0.1	&	0.3	&	0.5	&	0.7	&	1	\\\hline
	&	Intercept	&	0.9953	&	1.0248	&	1.0196	&	1.0192	&	1.0151	&	0.8395	\\	
Without	&		&	(0.17)	&	(0.17)	&	(0.18)	&	(0.19)	&	(0.21)	&	(0.26)	\\	
outliers	&	log(time)	&	3.0554	&	3.0294	&	3.0266	&	3.0287	&	3.0239	&	3.1739	\\	
	&		&	(0.15)	&	(0.15)	&	(0.16)	&	(0.17)	&	(0.19)	&	(0.23)	\\\hline
	&	Intercept	&	1.2431	&	1.1680	&	1.0630	&	1.2100	&	1.1587	&	1.0191	\\	
With one	&		&	(0.16)	&	(0.17)	&	(0.18)	&	(0.18)	&	(0.2)	&	(0.25)	\\	
outlier	&	log(time)	&	2.8404	&	2.9059	&	2.9955	&	2.8596	&	2.8993	&	3.0182	\\	
	&		&	(0.14)	&	(0.15)	&	(0.16)	&	(0.16)	&	(0.18)	&	(0.22)	\\	\hline
	&	Intercept	&	1.7291	&	1.2360	&	1.2496	&	1.0812	&	1.1722	&	1.0758	\\	
With two	&		&	(0.15)	&	(0.16)	&	(0.17)	&	(0.19)	&	(0.2)	&	(0.24)	\\	
outlier	&	log(time)	&	2.2968	&	2.8271	&	2.8153	&	2.9647	&	2.8805	&	2.9542	\\	
	&		&	(0.14)	&	(0.14)	&	(0.15)	&	(0.17)	&	(0.18)	&	(0.22)		\\	\hline
\end{tabular}
\end{center}
  \end{table}

Next we will create some artificial outliers in this ``good" data and check how the estimators 
are affected by those outliers for several $\alpha \geq 0$. We will consider two cases; --- 
case (i)  will have one outlier where we replace the first observation (at time 1) from 1 to 10, and 
case (ii) will have two outliers where we retain the  outlier of case (i) and also replace the last observation (at time 20) from 159 to 15. The MDPDE for different $\alpha \geq 0$ are shown in Table \ref{TAB:est_AIDS_Data} for both these cases of outliers. It is clear from the table that the 
MDPDE corresponding to $\alpha >0.3$ do not differ significantly after the insertion of the outliers; but the MLE (corresponding to $\alpha=0$) changes drastically. 

\subsection{Leukemia Data}

Our next data example consist of the observation on 33 leukemia patients obtained from Cook and Weisberg 
(1982, p. 193).
Here, we will model the data by the logistic regression with a binary response variable taking value one for at least 
52 survival of a leukemia patients and two covariates; --- ``WBC", the white blood cell count (in multiple of $10^4$) and ``AG", a binary indicator of the presence of Acute Granuloma in the white blood cells (Hosseinian, 2009). It was identified by Cook and Weisberg (1982) that the $15^{th}$  observation corresponds to a patients surviving for a long period despite of having large WBC counts of 100000 and 
so it generates an outlier in the data influencing the MLE.

	Here, we estimate the coefficients of the fitted logistic regression model robustly 
	avoiding the effect of the outlying observation in the data. For we apply the proposed 
	MDPDE for several values of $\alpha$ for all the data including the outlier ($15^{th}$  observation)
	 and also to the reduced data without the $15^{th}$  observation. The estimated parameter values
	 along with their asymptotic standard error are presented in Table \ref{TAB:est_leukemia_Wout} 
	 and \ref{TAB:est_leukemia_WOout} respectively. Comparing the two tables
	 we can see that all the MDPDE corresponding to $\alpha \geq 0.3$ are almost unaffected by the 
	 oulier observation and gives similar results to the MLE after excluding outlier.

  \begin{table}[h]
\caption{The minimum DPD estimates of the parameters for the Leukemia Data. The standard error of the estimates are given in the parenthesis.}
\label{TAB:est_leukemia_Wout}
\begin{center}
\begin{tabular}{lrrr} \hline
  $\alpha$	&	Intercept	&	AG	&	WBC 	\\\hline
0	&	-1.3059  (0.81)	&	2.2613  (0.95)	&	-0.3181  (0.19)	\\
0.1	&	-1.2426  (0.82)	&	2.2058  (0.97)	&	-0.3405  (0.2)	\\
0.3	&	0.1017  (1.15)	&	2.4381  (1.39)	&	-2.0017  (1.39)	\\
0.5	&	0.1386  (1.27)	&	2.4574  (1.62)	&	-2.0246  (1.7)	\\
0.7	&	0.1376  (1.38)	&	2.4512  (1.87)	&	-1.9844  (1.97)	\\
1	&	0.1442  (1.58)	&	2.459  (2.35)	&	-1.9635  (2.47)		\\	\hline
\end{tabular}
\end{center}
\end{table}

  \begin{table}[h]
\caption{The minimum DPD estimates of the parameters for the Leukemia Data (without the $15^{th}$  observation). The standard error of the estimates are given in the parenthesis.}
\label{TAB:est_leukemia_WOout}
\begin{center}    		
\begin{tabular}{lrrr} \hline
$\alpha$	&	Intercept	&	AG	&	WBC 	\\\hline
0	&	0.2152  (1.08)	&	2.5582  (1.24)	&	-2.3609  (1.36)	\\
0.1	&	0.1868  (1.10)	&	2.5261  (1.28)	&	-2.253  (0.14)	\\
0.3	&	0.1544  (1.17)	&	2.4826  (1.43)	&	-2.111  (1.48)	\\
0.5	&	0.1407  (1.27)	&	2.4592  (1.62)	&	-2.0286  (1.7)	\\
0.7	&	0.1374  (1.38)	&	2.4516  (1.87)	&	-1.9846  (1.97)	\\
1	&	0.1436  (1.58)	&	2.458  (2.35)	&	-1.9623  (2.46)		\\	\hline
   \end{tabular}
\end{center}
  \end{table}

\subsection{Skin Data}

Now, we will consider the popular skin data set of Finney (1947) on occurrence of 
``vaso constriction" in the skin of digits after air inspiration. The data set was obtained 
from a controlled study and analyzed by Pregibon (1982) and by Croux and Haesbroeck (2003). 
Again we will model this data by means of the logistic regression model where 
the binary response gives the occurrence of vaso constriction in the skin of digits after a single deep
 breath and the explanatory variables are the logarithm of the volume of air inspired (``log.Vol")
 and the logarithm of the inspiration rate (``log.Rate").

It can be seen 
that the $4^{th}$
and $18^{th}$ observations influence the MLE to make it difficult to partition the responses.
However, after deleting these two observations the overlap between the two outcome of the response 
variable depends only through on observation giving the facility to partition the outcomes 
with only one error. So, here also we will apply the proposed MDPDE for different $\alpha$
on the whole data including the influential observations as well as on the outlier-deleted 
data after dropping observation 4 and 18.  The results are presented in Table \ref{TAB:est_skin_Wout} 
and \ref{TAB:est_skin_WOout} respectively. Again, we can se from the tables that the MDPDEs 
corresponding to tunning parameter $\alpha \geq 0.3$ gives us robust estimators which remains 
unaffected by the two influential observations and generates the similar estimators as the most
 efficient MLE for the outlier free data.

  \begin{table}[h]
\caption{The minimum DPD estimates of the parameters for the Skin Data. The standard error of the estimates are given in the parenthesis.}
  \label{TAB:est_skin_Wout}
\begin{center}
\begin{tabular}{lrrr} \hline
$\alpha$	&	Intercept	&	log(Rate)	&	log(Vol)	\\\hline
0	&	-2.88  (1.32)	&	4.56  (1.84)	&	5.18  (1.86)	\\
0.1	&	-3.14  (1.48)	&	4.83  (2.07)	&	5.46  (2.12)	\\
0.3	&	-19.05  (12.89)	&	24.89  (16.65)	&	30.57  (21.68)	\\
0.5	&	-21.05  (18.06)	&	27.44  (23.42)	&	34.13  (30.54)	\\
0.7	&	-20.85  (21.35)	&	27.2  (27.86)	&	33.97  (36.27)	\\
1	&	-23.77  (32.98)	&	31.08  (43.45)	&	39.34  (56.35)		\\	\hline
\end{tabular}
\end{center}
\end{table}

  \begin{table}[h]
\caption{The minimum DPD estimates of the parameters for the Skin Data (without the $4^{th}$ and $18^{th}$ observations). The standard error of the estimates are given in the parenthesis.}
  \label{TAB:est_skin_WOout}
\begin{center}
\begin{tabular}{lrrr} \hline
$\alpha$	&	Intercept	&	log(Rate)	&	log(Vol)	\\\hline
0	&	-24.58  (14.02)	&	31.94  (17.76)	&	39.55  (23.25)	\\
0.1	&	-24.13  (14.89)	&	31.36  (18.97)	&	38.9  (24.81)	\\
0.3	&	-22.01  (15.86)	&	28.66  (20.42)	&	35.57  (26.66)	\\
0.5	&	-21.14  (18.16)	&	27.55  (23.55)	&	34.28  (30.71)	\\
0.7	&	-20.86  (21.43)	&	27.21  (27.94)	&	33.97  (36.37)	\\
1	&	-23.77  (32.98)	&	31.08  (43.45)	&	39.34  (56.35)		\\	\hline
   \end{tabular}
\end{center}  
\end{table}

\subsection{Damaged Carrots Data}

As an interesting data example leading to the logistic regression, 
we will consider the damaged carrots dataset of Phelps (1982). 
The data set was obtained from a soil experiment trial containing the proportion of insect damaged
 carrots with three blocks and eight dose levels of insecticide in the experiments and discussed by
Williams (1987). McCullagh and Nelder (1989) this data to 
illustrate the identification methods for isolated departures from the model through an outlier in 
the y-space present in the data ($14^{\rm th}$ observation; dose level 6 and block 2). 
Later Cantoni and Ronchetti (2001) modeled this data by a binomial logistic model to 
illustrate the performance of their proposed robust estimators. 
However, it can be checked easily that the observation 14 is only  the outlier in y-space and not a leverage point.

 We will now apply the Minimum density power divergence estimation for several different $\alpha$ to see the performance 
 of the proposed method in case of presence of outlier only in y-space. Table \ref{TAB:est_carrot_data}
 presents the parameter estimates, their asymptotic standard error and corresponding p-value for 
 different tuning parameter $\alpha$. The estimators corresponding to $\alpha \ge 0.3$ again turns out 
 to be highly robust and also similar to the robust estimator obtained by Cantoni and Ronchetti (2001). 
 Also,  for these estimators the indicator of Block 1 turns out to be insignificant which became 
 significant in case of the maximum likelihood estimator (corresponding to $\alpha=0$) due to the 
 outlying observation.

 \begin{table}[h]
\caption{The minimum density power divergence estimates, their standard errors and p-values for the Damage Carrots Data}
 \label{TAB:est_carrot_data}
 \begin{center}
\begin{tabular}{llrrrrrr} \hline
		&		&	\multicolumn{1}{c}{$\alpha=0$}	&	\multicolumn{1}{c}{$\alpha=0.1$}	&	\multicolumn{1}{c}{$\alpha=0.3$}	&	\multicolumn{1}{c}{$\alpha=0.5$}	&	\multicolumn{1}{c}{$\alpha=0.7$}	&	\multicolumn{1}{c}{$\alpha=1$}	\\ \hline
	Intercept	&	Estimate	&	1.4805	&	1.4880	&	1.4974	&	1.5157	&	1.5310	&	1.5569	\\
		&	SE 	&	0.6562	&	0.6648	&	0.6859	&	0.7118	&	0.7406	&	0.7868	\\
		&	P-Value	&	0.0339	&	0.0352	&	0.0395	&	0.0441	&	0.0501	&	0.0599	\\ \hline
	logdose	&	Estimate	&	-1.8175	&	-1.8163	&	-1.8102	&	-1.8102	&	-1.8102	&	-1.8152	\\
		&	SE 	&	0.3439	&	0.3484	&	0.3601	&	0.3749	&	0.3917	&	0.4183	\\
		&	P-Value	&	0.0000	&	0.0000	&	0.0000	&	0.0001	&	0.0001	&	0.0002	\\ \hline
	Block1	&	Estimate	&	0.5421	&	0.5330	&	0.5149	&	0.4969	&	0.4824	&	0.4654	\\
		&	SE 	&	0.2318	&	0.2338	&	0.2392	&	0.2462	&	0.2542	&	0.2668	\\
		&	P-Value	&	0.0284	&	0.0322	&	0.0421	&	0.0554	&	0.0704	&	0.0945	\\ \hline
	Block2	&	Estimate	&	0.8430	&	0.8284	&	0.7973	&	0.7710	&	0.7483	&	0.7240	\\
		&	SE 	&	0.2260	&	0.2283	&	0.2344	&	0.2422	&	0.2510	&	0.2649	\\
		&	P-Value	&	0.0011	&	0.0014	&	0.0025	&	0.0041	&	0.0067	&	0.0119 	\\\hline
 \end{tabular}
 \end{center}
 \end{table}

\section{Conclusion}\label{SEC:conclusion}

In this paper, we have proposed a new methodology for robust estimation in case of generalised 
linear models and considered two prominent special cases -- Poisson regression and logistic regression. 
These models are very useful for analyzing count response and binary response respectively.
We have established the robustness properties of the proposed method in terms of the influence function
analysis and also applied it to several real data sets having different types of outliers. 
In these we have seen that the proposed estimators with moderate $\alpha >0$ are highly robust in handling 
different kinds of outliers and generate robust estimates competitive to the existing ones in the literature. 
In fact the estimated standard errors of our parameter estimates are smaller 
than those of the existing robust estimates in most cases for the real data examples studied by us 
for all moderately small $\alpha\leq 0.5$. In some cases the same holds for $\alpha$ as high as $0.7$. 
Thus we hope that the estimators discussed in this paper will help the researchers in several fields 
such as medical biology, epidemiology and controlled trial experiments in biological sciences 
to estimate the model parameters in generalised linear model (including Poisson and logistic regression) 
efficiently and robustly even in contaminated scenarios.
%

 \begin{table}[h] 
 	\def~{\hphantom{0}}
\caption{The Optimum values of the tunning parameter $\alpha$ for different data sets obtained using several ``Pilot" values of $\alpha$}
 \label{TAB:opt_alpha}
\begin{center} 	  		
\begin{tabular}{lrrrrrr} \hline
& \multicolumn{6}{c}{Pilot $\alpha$} \\
Data Sets	&	0	&	0.1	&	0.3	&	0.5	&	0.7	&	1	\\\hline
AIDS Data	&	0	&	0.05	&	0.05	&	0.1	&	0.1	&	1	\\
AIDS Data	&	0	&	0.2	&	0.65	&	0.35	&	0.55	&	0.45	\\
(with one outlier)	&		&		&		&		&		&		\\
AIDS Data	&	0	&	0.3	&	0.3	&	0.55	&	0.5	&	0.55	\\
(with two outliers)	&		&		&		&		&		&		\\
Epilepsy Data	&	0	&	0.05	&	0.35	&	0.3	&	1	&	0.95	\\
Leukemia Data	&	0	&	0.1	&	0.3	&	0.3	&	0.3	&	0.3	\\
Leukemia Data	&	0	&	0	&	0.1	&	0.1	&	0.1	&	0.1	\\
(without outlier)	&		&		&		&		&		&		\\
Skin Data	&	0	&	0.1	&	0.3	&	0.35	&	0.35	&	0.4	\\
Skin Data	&	0	&	0	&	0.25	&	0.3	&	0.35	&	0.05	\\
(without outlier)	&		&		&		&		&		&		\\
Carrots Data	&	0	&	0.05	&	0.3	&	0.55	&	0.7	&	0.95		\\	\hline
 \end{tabular}
\end{center} 
\end{table}


%

\label{lastpage}


\begin{thebibliography}{}

\bibitem{bhhj98}
Basu, A., Harris, I. R., Hjort, N. L. and Jones, M. C. (1998). Robust and efficient estimation by minimizing a density power divergence. {\it Biometrika}, {\bf 85}, 549--559.

\bibitem{bsp11}
Basu, A., Shioya, H. and Park, C. (2011). {\it Statistical Inference: The Minimum Distance Approach}. Chapman $\&$ Hall/CRC, Boca Raton, Florida.

\bibitem{cr01}
Cantoni, E. and Ronchetti, E. (2001). Robust inference for generalised
linear models. {\it J. Amer. Statist. Assoc.}, {\bf 96(455)}, 1022–-1030.

\bibitem{cw82}
Cook, R. D. and Weisberg, S. (1982). Residuals and influence in regression.
{\it Monographs on Statistics and Applied Probability}. Chapman $\&$ Hall, London.

\bibitem{ch03}
Croux, C. and Haesbroeck, G. (2003). Implementing the Bianco and
Yohai estimator for logistic regression. {\it Comput. Statist. Data Anal.}, {\bf 44(1-2)}, 273-–295. Special issue in honour of Stan Azen: a birthday celebration.

\bibitem{d02}
Dobson, A. J. (2002). {\it An introduction to generalised linear models}. Second
edition. Chapman $\&$ Hall/CRC Texts in Statistical Science Series.
Chapman $\&$ Hall/CRC, Boca Raton, Florida.

\bibitem{f47}
Finney, D. J. (1947) The estimation from individual records of relationship
between dose and quantal response. {\it Biometrika} {\bf 34(3/4)}, 320–-334.


\bibitem{gb13}
Ghosh, A. and Basu, A. (2013a). Robust Estimation for Independent Non-Homogeneous Observations using Density Power Divergence with Applications to Linear Regression. {\it Electronic Journal of statistics}, {\bf 7}, 2420--2456.

\bibitem{gb14}
Ghosh, A. and Basu, A. (2013b). Robust Estimation for Non-Homogeneous Data and the Selection of the Optimal Tuning Parameter: The Density Power Divergence Approach. {\it Under Review}


\bibitem{hk01}
Hong, C. and Kim, Y. (2001). Automatic selection of the tuning parameter in the minimum density power
divergence estimation. {\it Journal of the Korean Statistical Society}, {\bf 30}, 453--465.


\bibitem{h09}
Hosseinian, S. (2009).  {\it Robust Inference for Generalized Linear Models:
Binary and Poisson Regression}. Ph. D. Thesis. Ecole Polytechnique Fédéral de Lausanne, CH-1015, Lausanne.
 
\bibitem{l85}
Leppik, I.~E., et al.~(1985) A double-blind crossover evaluation of progabide in partial seizures. 
{\it Neurology}, {\bf 35}, 285.

 
\bibitem{mn83}
McCullagh, P. and Nelder, J. A. (1989) {\it Generalized linear models (Second Edition)}.
Monographs on Statistics and Applied Probability. Chapman $\&$ Hall, London.

\bibitem{p82}
Phelps, K. (1982), Use of the Complementary log-log Function to Describe Dose-response Relationships in Insecticide Evaluation Field Trials, in Lec- ture Notes in Statistics, 14. GLIM.82: {\it Proceedings of the International Conference on Generalized Linear Models}, ed. R. Gilchrist, New York: Springer-Verlag.


\bibitem{p82}
Pregibon, D. (1982). Resistant fits for some commonly used logistic
models with medical applications. {\it Biometrics}, {\bf 38(2)}, 485-–498.



\bibitem{tv90}
Thall, P. F. and Vail, S. C. (1990) Some covariance models for longitudinal
count data with overdispersion. {\it Biometrics}, {\bf 46(3)},  657-–671.


\bibitem{wj05}
Warwick, J. and Jones, M. C. (2005) Choosing a robustness tuning parameter, {\it Journal of Statistical
Computation and Simulation}, {\bf 75}, 581--588.

\bibitem{w87}
Williams, D. A. (1987), generalised linear model Diagnostics Using the Deviance and Single Case Deletions, {\it Applied Statistics}, {\bf 36}, 181--191.

\end{thebibliography}
\end{document}